\documentclass[a4paper]{article}
\usepackage{amsmath}
\usepackage{amssymb}
\usepackage{epsfig}
\usepackage{graphicx}
\usepackage{natbib}
\usepackage{subfigure}
\usepackage{fullpage}

\newcommand{\kms}{km s$^{-1}$}
\newcommand{\gcc}{g cm$^{-3}$}
\newcommand{\be}[1]{\begin{equation}\label{#1}}
\newcommand{\ee}{\end{equation}}

\def\araa{ARA\&A}%
\def\apj{ApJ}%
\def\apjl{ApJ}%
\def\apjs{ApJS}%
\def\ao{Appl.~Opt.}%
\def\aap{A\&A}%
\def\solphys{Sol.~Phys.}%
\def\nat{Nature}%
\begin{document}

\author{M.~C.~M. Cheung\footnote{\emph{Present address:} Lockheed Martin Solar and Astrophysics
Laboratory, Bldg/252 3251 Hanover St., Palo Alto, CA 94304, USA.}, M. Sch\"ussler \\
Max Planck Institute for Solar System Research\\
37191 Katlenburg-Lindau, Germany\\
e-mail: [cheung,msch]@mps.mpg.de \\
\and F. Moreno-Insertis \\
Instituto de Astrof\'{i}sica de Canarias\\
38200 La Laguna (Tenerife), Spain \\
Dept of Astrophysics, Faculty of Physics
\\University of La Laguna, 38200 La Laguna (Tenerife), Spain \\
e-mail: fmi@ll.iac.es}

\title{Magnetic flux emergence in granular convection: \\Radiative MHD simulations and observational signatures}
\date{19 February 2007} \maketitle

\abstract{We study the emergence of magnetic flux from the
near-surface layers of the solar convection zone into the
photosphere. To model magnetic flux emergence, we carried out a
set of numerical radiative magnetohydrodynamics simulations. Our
simulations take into account the effects of compressibility,
energy exchange via radiative transfer, and partial ionization in
the equation of state. All these physical ingredients are
essential for a proper treatment of the problem. Furthermore, the
inclusion of radiative transfer allows us to directly compare the
simulation results with actual observations of emerging flux. We
find that the interaction between the magnetic flux tube and the
external flow field has an important influence on the emergent
morphology of the magnetic field. Depending on the initial
properties of the flux tube (e.g. field strength, twist, entropy
etc.), the emergence process can also modify the local granulation
pattern. The emergence of magnetic flux tubes with a flux of
$10^{19}$ Mx disturbs the granulation and leads to the transient
appearance of a dark lane, which is coincident with upflowing
material. These results are consistent with observed properties of
emerging magnetic flux.}

\section{Introduction}
\label{sec:introduction} Solar magnetic fields on the surface of
the Sun exist and evolve over a wide range of length- and
time-scales. The most prominent magnetic features on the solar
surface are sunspots. In addition, there is a whole hierarchy of
magnetic features including pores, micropores, plages and faculae.
An~\emph{active region} is an extended bipolar configuration on
the solar surface resulting from the emergence of magnetic fields
from the convection zone~\citep{Parker:FormationOfSunspots}. In
terms of the amount of magnetic flux in each polarity, there is a
continuous spectrum of active region
sizes~\citep{HagenaarSchrijver:BipolarRegions}. Although a
partitioning of the flux spectrum for the sake of classification
may seem somewhat arbitrary, it allows us to conveniently refer to
active regions of different sizes.~\emph{Large active regions}
have polarities containing a flux of $>5\times 10^{21}$~Mx and
contain sunspots. In large active regions, the magnetic flux is
shared amongst the whole spectrum of magnetic features. Large
active regions have lifetimes of up to months.~\emph{Small active
regions}, which contain a flux of $1\times 10^{20} - 5\times
10^{21}$ Mx in each polarity, may consist of pores and smaller
magnetic features but lack sunspots. Small active regions may
persist for up to days to weeks.~\emph{Ephemeral active regions}
have even less flux ($3\times 10^{18} - 1\times 10^{20}$ Mx), and
have lifetimes of only hours to days
\citep{Zwaan:ElementsAndPatterns}. Often, ephemeral active regions
are simply referred to as ephemeral regions.

The characteristic timescale for the emergence of flux is
typically only a small fraction of the lifetime of an active
region. For instance, almost all the flux of a large active region
emerges within the first $\sim 4$ days of their
development~\citep{Zwaan:EmergenceOfMagneticFlux,Hagenaar:Ephemeralregions}.
Observations of the birth of an active region (AR) indicate that
the total flux in each polarity of the AR is not the consequence
of the emergence of a coherent, monolithic flux bundle. Rather, it
builds up as the result of many small flux bundles emerging
simultaneously or in succession. An Emerging Flux Region (EFR) is
the area on the solar surface where these emergence events take
place~\citep{Zirin:FineStructureOfSolarMagneticFields,Zwaan:EmergenceOfMagneticFlux}.
The onset of the birth of an active region is characterized by the
appearance of a compact and very bright plage. The plage consists
of magnetic flux elements of opposite polarity, which move apart
at an initial velocity of about $2$~\kms. New flux continues to
emerge near the polarity inversion line. The orientation of the
emerging field is not random. Rather, it is roughly aligned along
the axis connecting the two polarities. As a result of emergence,
flux is accumulated in both polarities. If sufficient flux has
emerged, pores and possibly sunspots appear. These tend to be
formed near the leading and following edges of the expanding
plage~\citep[see~][~and references
therein]{Zwaan:EmergenceOfMagneticFlux}.

The granulation pattern in an EFR may appear different than that
of the quiet Sun. In quiet-Sun granulation, bright granules
correspond to upflows whereas the dark intergranular boundaries
consist of downflow lanes and vertices. Transient dark alignments
in the central part of an EFR have been detected in continuum
images and also in maps of the core intensity of photospheric
spectral lines~\citep*{BrayLoughhead:SolarGranulation,
BrantsSteenbeek:EFR,Zwaan:EmergenceOfMagneticFlux,
StrousZwaan:SmallScaleStructure}. The darkenings are roughly
aligned along the axis connecting the two polarities of the active
region and typically last about $10$ minutes. In the continuum,
they are darker than the intergranular boundaries but the spectral
lines show Doppler shifts corresponding to upward velocities
exceeding $0.5$~\kms. By estimating the diameter and strength of
emerging flux bundles that lead to the appearance of dark
alignments in an EFR,~\citet{BrantsSteenbeek:EFR} estimated that
each bundle contains a longitudinal flux of about $10^{19}$ Mx.
This value is consistent with the estimate given
by~\citet{Born:YoungActiveRegions}, who measured the total flux of
an active region and counted the number of arch filaments in
H$\alpha$.

On the basis of the observations previously
summarized,~\citet{Zwaan:AppearanceOfMagneticFlux,Zwaan:EmergenceOfMagneticFlux}
constructed a heuristic model of an EFR at and below the solar
surface. In his scenario, flux emerges as a collection of arched
flux tubes rising through the convection zone. Deep in the
convection zone, the flux tubes connect to an underlying coherent
large flux tube. Near the surface, however, they are separated
from each other. In his model, the transient dark alignments in
the observations are locations of the apices of arched flux tubes
emerging at the surface, which appear dark because the horizontal
magnetic field suppresses turbulent heat exchange. After the
horizontal top of a tube has emerged, the photospheric footpoints
of the tubes separate and the field at the footpoints become
increasingly vertical. The coalescence of the vertical flux
elements in each polarity may lead to the formation of pores and
sunspots if sufficient flux is accumulated.

Subsequent observations of EFRs are consistent with various
aspects of Zwaan's heuristic
model.~\citet{StrousZwaan:SmallScaleStructure} performed a
statistical analysis of over two hundred emergence events in a
single EFR. They found that emerging flux is characterized by the
transient appearance of dark alignments between the polarities.
Often, they observed the appearance of faculae at the ends of the
dark alignments. The faculae are typically associated with
downflows. More recently, ~\citet{Lites:EmergingFieldsVector}
and~\citet{Kubo:EmergingFluxRegion} have carried out observations
of the full Stokes vector in emerging flux regions and report the
detection of horizontal fields with strengths of about
$200-600$~G. As the newly emerged flux moves away from the
emergence site, the fields become vertical and obtain strengths in
excess of $1000$~G.

The aim of this study is to understand the process of magnetic
flux emergence from the convection zone into the photosphere by
means of numerical simulations. At present, it is not
computationally feasible to carry out numerical simulations that
include all the relevant physics for the emergence of magnetic
flux through all layers of the solar convection zone and
atmosphere. To progress, different aspects of the problem have to
be addressed separately.

One line of previous research has focused on studying the rise of
buoyant magnetic flux tubes in a stratified layer by means of 2D
and 3D MHD simulations. Most of these simulations~\citep[see, for
example,][]{Schussler:buoyancyrevisited,Moreno-InsertisEmonet:RiseOfTwistedTubes,LongcopeFisherArendt,EmonetMoreno-Insertis:PhysicsOfTwistedFluxTubes,Fan:2DTubes,
Cheung:MovingMagneticFluxTubes} start from a buoyant magnetic flux
tube embedded in an initially static atmosphere.
\citet{Dorch:3DTubeConvection} was the first to study the
interaction of magnetic flux tubes with convective flows. They
found that the convective motion of the flow external to the tube
enhances the loss rate of flux from the
tube.~\citet*{Fan:3DTubeConvection}
and~\citet*{Abbett:3DTubeInConvection} have also studied the
interaction of magnetic flux tubes and convective flow. They found
that flux tubes with field strengths exceeding the equipartition
value with respect to the kinetic energy density of the convective
motions are able to resist deformation by the convective flow,
while significantly weaker flux tubes are severely distorted.

Another line of work has focused on studying the dynamics of
toroidal magnetic flux tubes rising in the solar convection zone.
Numerical simulations of rising toroidal flux tubes under
the~\emph{thin flux tube
approximation}~\citep{RobertsWebb:VerticalMotions,Spruit:ThinFluxTube}
have been successful at reproducing the large-scale properties of
active regions, such as their emergence latitudes, their tilt
angles as well as the asymmetries between leading and following
polarities~\citep{DSilva:ARTiltAngle,Fan:ThinFluxTubeI,Fan:ThinFluxTubeII,
Caligari:EmergingTubesPartI,Caligari:EmergingTubesPartII}. When
the apices of magnetic loops reach a depth of about $10-20$ Mm
below the photosphere, their cross-sections have expanded so much
that the thin flux tube approximation no longer applies. This
limitation means that fully multi-dimensional MHD simulations must
be carried out in order to study the details of the emergence
process.

Beginning with the work of~\citet*{Forbes:FluxEmergence}
and~\citet{Shibata:FluxEmergence}, a variety of flux emergence
simulations have been carried out. The focus of most studies in
the literature has been on the evolution of the emerging magnetic
field in the
corona~\citep{Fan:2001_TubeEmergence,Magara:2.5D,MagaraLongcope:InjectionMagneticEnergyHelicity,Fan:3DCoronalFields,
Abbett:Emergence,Archontis:EmergenceIntoCorona,
Archontis:EmergenceIntoCoronaII,TorokKliem:KinkUnstableFluxRopes,
Isobe:Rayleigh-TaylorInstability,Galsgaard:FluxEmergenceReconnection,LeakeArber:EmergenceThroughPartiallyIonized,
Murray:EmergingFluxTube,Galsgaard:TBD}.
Recently,~\citet{Magara:2006} carried out MHD simulations of flux
emergence to study the evolution of the coronal and photospheric
field of an emerging bipolar region. Like the aforementioned
references, his work is relevant for the large scale behaviour of
active regions and ignores the effects of convection as well as
radiative energy exchange.

In this study, we examine in detail the effects of convective
flows and radiative transfer on emerging magnetic flux. To this
end, we restrict our attention to flux emergence from the
near-surface part of the convection zone into the photosphere. The
article is structured as follows.
In~Sect.~\ref{sec:equations_methods}, we present the system of
radiative MHD equations, the numerical methods used to solve this
system as well as the setup of the simulations.
In~Sect.~\ref{sec:subsurface_dynamics}, we discuss the dynamics of
rising flux tubes with special emphasis on their interaction with
the convective motions.
In~Sect.~\ref{sec:observational_signatures}, we discuss the
observational signatures of emerging flux as obtained from our
simulations and compare them with observational studies in the
literature. Finally, we close the paper with a discussion
in~Sect.~\ref{sec:summary}.

\section{Governing equations, numerical methods and simulation setup}
\label{sec:equations_methods}

\subsection{Governing equations}
The system of radiative MHD equations (in Gaussian cgs units)
consists of
\begin{eqnarray}
\frac{\partial \varrho}{\partial t} &+& \nabla \cdot (\varrho
\vec{v})=0\label{eqn:continuity}, \\
\frac{\partial \varrho \vec{v}}{\partial t} &+& \nabla \cdot
\left[ \varrho \vec{v}\otimes\vec{v} + p_{\rm tot}
\underline{\underline{1}} -
\frac{\vec{B}\otimes\vec{B}}{4\pi}\right ] = \varrho \vec{g} + \nabla \cdot \underline{\underline{\tau}},\label{eqn:momentum}\\
\frac{\partial e}{\partial t} &+& \nabla\cdot \left [
\vec{v}(e+p_{\rm tot}) -
\frac{1}{4\pi}\vec{B}(\vec{v}\cdot\vec{B}) \right ] = \varrho (\vec{g}\cdot\vec{v}) \label{eqn:energy}\\
+Q_{\rm rad}&+&
\frac{1}{4\pi}\nabla\cdot(\vec{B}\times\eta\nabla\times\vec{B}) +
\nabla\cdot(\vec{v}\cdot\underline{\underline{\tau}}) +
\nabla\cdot(K\nabla T) \nonumber \\
\frac{\partial \vec{B}}{\partial t} &+& \nabla \cdot
\left[\vec{v}\otimes\vec{B} - \vec{B}\otimes\vec{v} \right] = -
\nabla \times(\eta\nabla\times\vec{B}),\label{eqn:induction}
\end{eqnarray}

\noindent where $\varrho$ is the fluid mass density, $\vec{v}$ the
fluid velocity, $\vec{B}$ the magnetic field, $p_{\rm tot}=
p+|\vec{B}|^2/8\pi$ the sum of the gas and magnetic pressures, and
$e=\varrho\varepsilon + \frac{1}{2}\rho |\vec{v}|^2
+|\vec{B}|^2/8\pi$ the sum of the internal, kinetic and magnetic
energy densities per unit volume. The symbol $\otimes$ denotes a
tensor product and $\underline{\underline{1}}$ is the $3\times3$
identity matrix. $T$ is the gas temperature and $K$ the thermal
conductivity. $\underline{\underline{\tau}}$ represents the
viscous stress tensor. $Q_{\rm rad}$ is the radiative heating rate
and $\vec{g}=-2.74\times10^4\hat{z} $ cm s$^{-2}$ is the
gravitational acceleration.

Equations (\ref{eqn:continuity}), (\ref{eqn:momentum}) and
(\ref{eqn:energy}) express, respectively, the principles of mass
conservation, momentum balance and energy conservation, while
Eq.~(\ref{eqn:induction}) is the induction equation, which
describes the evolution of a magnetic field in the plasma. In this
equation, $\eta=c^2/4\pi\sigma$ is the magnetic diffusivity, $c$
the speed of light and $\sigma$ the electrical conductivity.

The MURaM code was used to carry out the radiative MHD simulations
in this study. The simulations presented here used grey radiative
transfer (i.e. 1 frequency bin). For details about the numerical
methods implemented in MURaM, we refer the reader
to~\citet{Voegler:PhD} and~\citet{Voegler:MURaM}.

\subsection{Boundary conditions}
\label{subsec:two_d_bcs} Periodic boundary conditions are imposed
at the vertical boundaries. The open boundary condition at the
lower boundary is identical to the one used
by~\citet{Voegler:MURaM}.

The top boundary condition has been modified to allow for the
advection of magnetic field through the top boundary. This is
important because we do not want the emerged magnetic field to be
artificially trapped in the photosphere. In principle, the
following conditions
\begin{equation}
\frac{\partial v_x}{\partial z} = \frac{\partial v_y}{\partial z}
= \frac{\partial v_z}{\partial z} =
0,\label{eqn:stress_free_conditions}
\end{equation}
\noindent should suffice for a smooth outflow. This condition,
however, is independent of whether the mass flux through the upper
boundary is appropriate or unrealistically high. In order to keep
the mass flux at appropriate levels, additional constraints must
be applied. To this end, we
follow~\citet*{SteinNordlund:SolarGranulation} and implement
a~\emph{fiducial layer} above the top boundary of our simulation
domain.

The magnetic field above the upper boundary of the domain is
matched to a potential field. This requires an extrapolation of
the magnetic field in the uppermost domain layer into the ghost
cells at each time step. For further details, we refer the reader
to~\citet{Cheung:PhD}.

\subsection{Initial conditions}

\begin{table*}
\centering
\begin{tabular}{ccccccc}
  \hline
  Label & $B_0$(G) & $\lambda$ & $\Phi_0$ ($10^{19}$ Mx) & $\Phi_{\rm max}^{z=0}$ ($10^{19}$ Mx) &Emergence fraction (\%)\\
  \hline
  U1 & $8500$ & $0.5$ & $1$ &$7.7$ & $67$\\
  U2 & $8500$ & $0.25$ & $1$ &$4.7$ & $68$\\
  U3 & $8500$ & $0.1$ & $1$ &$3.1$ & $44$\\
  U4 & $8500$ & $0$ & $1$ &$2.6$ & $30$\\
  U5 & $8500$ & $0.75$ & $1$ &$11$ & $62$\\
  U6 & $2500$ & $0.5$ & $0.31$ &$0.9$ & $12$\\
  U7 & $5250$ & $0.5$ & $0.66$ &$3.0$ & $39$\\
  U8 & $7000$ & $0.5$ & $0.86$ &$5.6$ & $49$\\
  \hline
\end{tabular}
\caption{Initial properties of the individual magnetic flux tubes
and emerged flux of the simulation runs. The columns give (from
left to right) the label of the run, the magnitude of the
longitudinal field at the tube axis ($B_0$), the twist parameter
($\lambda$), the total longitudinal flux ($\Phi_0$), the
maximum~\emph{unsigned} magnetic flux ($\Phi_{\rm max}^{z=0}$)
crossing the $z=0$ plane (i.e. base of the photosphere) within a
time period of 20~min after the start of the simulation, and the
percentage of tracer fluid elements in the initial flux tube that
manage to rise to the photosphere ($z=0$) within the same time
period. In all cases, the characteristic radius of the initial
flux tube is $R_0=200$ km.}\label{table:simulation_runs}
\end{table*}

We have carried out a number of simulation runs, each modelling
the emergence of an individual flux tube initially embedded in the
near-surface layers of the convection zone. In order to prepare
the initial condition for the flux emergence simulations, a 3D
dynamic model of the near-surface layers of the convection zone
and photosphere is required. We obtained such a model by means of
a non-magnetic simulation with the MURaM code. The details of the
corresponding simulation setup are given in
\citet*{Cheung:ReversedGranulation2007}.

The simulation domain of our model atmosphere has dimensions
$24\times 12\times 2.3$ Mm$^3$. The level $z=0$ corresponds to the
mean geometrical height where the continuum optical depth at
$5000$ \AA~is unity (i.e. $\tau_{\rm 5000} = 1$) and is located at
a height of $1.85$ Mm above the bottom boundary of the domain. We
implant a horizontal, twisted magnetic flux tube with the axis of
the tube located at $[y,z] =[6,-1.35]$~Mm into a snapshot of the
3D hydrodynamic simulation. In all subsequent discussion, $t=0$
refers to the moment when the flux tube is introduced. The
longitudinal and transverse (azimuthal) components of the magnetic
field have the form
\begin{eqnarray}
B_l(r) & = & B_0 \exp{(-r^2/R_0^2)}, \label{eqn:long_profile}\\
B_\theta(r) & = & \frac{\lambda
r}{R_0}B_l,\label{eqn:transverse_profile}
\end{eqnarray}
where $r\in[0,2R_0]$ is the radial distance from the tube axis.
The longitudinal flux carried by the tube is $\Phi_0 = \int B_l
{\rm d}S\simeq 0.98\pi R_0^2 B_0$, so that $R_0$ can be viewed as
a characteristic `tube radius'. The dimensionless twist parameter
$\lambda$ specifies the relative strength of the transverse
component of the magnetic field with respect to the longitudinal
component.

Let $\vec{v}_{\rm orig}$ be the original velocity distribution in
the non-magnetic domain. In order to start with an almost
stationary flux tube without excessively perturbing the convective
flow, we impose the following initial velocity distribution within
the tube: $\vec{v}_{\rm tube}(r) = (1-e^{-r^2/R_0^2})\vec{v}_{\rm
orig}$. With increasing radial distance from the axis, the fluid
velocity approaches the original velocity distribution of the
external convecting flow.

For a given magnetic field distribution in the tube, the internal
gas pressure distribution is specified by requiring that the
divergence of the combined stress tensor (gas pressure $+$
Maxwell) be identical to that in the non-magnetic domain (the
divergence of the viscous stress tensor is not identical because
we have modified the velocity distribution in the magnetic flux
tube). Having done so, one is still free to choose the
distribution in the tube of one of the following thermodynamic
properties: mass density, specific entropy, or temperature. In all
of the simulations presented in this paper, the initial flux tube
was imparted with a uniform specific entropy distribution ($r\le
2R_0$). The value of the specific entropy was chosen as its
average value in the convective upflows at the initial depth of
the tube.

\section{Dynamics of subsurface rise and appearance at the photosphere}
\label{sec:subsurface_dynamics}
\subsection{The rise of a buoyant magnetic flux tube in the presence of convective flows}
\label{subsec:convection_flux_emergence} Convection plays an
important role in determining the properties of emerging flux.
Near-surface convection in the Sun consists of disjointed
upwellings separated by a network of downflow lanes and
vertices~\citep{SteinNordlund:Topology}. An initially horizontally
flux tube embedded in this setting will encounter both upflows and
downflows. Whereas the upwellings aid the emergence of some
segments of the tube, the downflows impede the rise of other
parts. Given such an initial configuration, under what
circumstances can we expect the tube to emerge coherently without
severe distortion by the downflows? In other words, what does it
mean to have a `strong' flux tube?

We expect that the flux tube can resist severe distortion by the
flow if the field strength satisfies $B_0 \gtrsim B_{\rm eq}$,
where $B_{\rm eq}$ is the field strength at which the magnetic
energy density and the kinetic energy density of the external flow
are in equipartition. The value of the equipartition field
strength depends on depth. At the original location of the flux
tube ($z=-1.35$ Mm), the ambient density is $\varrho = 4.2\times
10^{-6}$~\gcc. The typical vertical velocity of downflowing
material at this depth is $4-8$~\kms. For this range of
velocities, the corresponding range of equipartition field
strengths is $B_{\rm eq} = 2900 - 5800$ G. Taking the typical
values at the surface ($\varrho = 2.6\times10^{-7}$~\gcc, $v =
2-4$~\kms), the equipartition field strength there has much lower
values of $B_{\rm eq} = 450 - 700$ G.

Consideration of the force balance on the tube allows us to
determine a related, but more stringent criterion. Suppose that
the flux tube has uniform magnetic field strength, $B_0$, and
uniform density deficit, $\Delta \varrho$, with respect to the
surroundings. The buoyancy force experienced by the flux tube
overcomes the drag force exerted by a downflow of speed $v$ given
that~\citep[see, for
example,][]{Parker:SunspotsIIDrag,Moreno-Insertis:Risetimes,
Schuessler:1984, Schuessler:1987, Fan:3DTubeConvection}

\begin{equation}
B_0 \gtrsim \left( \frac{2 C_D \gamma_1}{\pi} \right)^{1/2} \left(
\frac{H_p}{R_0} \right)^{1/2} B_{\rm eq},
\label{ineq:force_balance2}
\end{equation}
\noindent where $H_p$ is the local pressure scale height, $C_D$
the drag coefficient (of order unity) and $\gamma_1$ the first
adiabatic exponent. The factor $2 C_D \gamma_1/\pi$ is of order
unity. The radius is important because the drag force (per unit
length) on the flux tube is proportional to the tube radius while
the buoyancy force (also per unit length) is proportional to the
square of the tube radius and to the square of the field strength.
This means that, for a given field strength, a thick tube tends to
be dominated by buoyancy while a sufficiently thin tube is
passively dragged by the surrounding flows. For example, we see
from the criterion given by Eq.~(\ref{ineq:force_balance2}) that a
tube with a radius $R_0 = 0.25 H_p$ must have a field strength of
at least $2 B_{\rm eq}$ in order to have sufficient buoyancy to
rise against the drag of the downflows. Since the downflows are
localized, the magnetic curvature of a deformed flux tube also
acts against the drag force. Like the buoyancy force, the
curvature force is also proportional to the square of the magnetic
field strength.

\begin{figure*}
\centering
\includegraphics[width=0.8\textwidth]{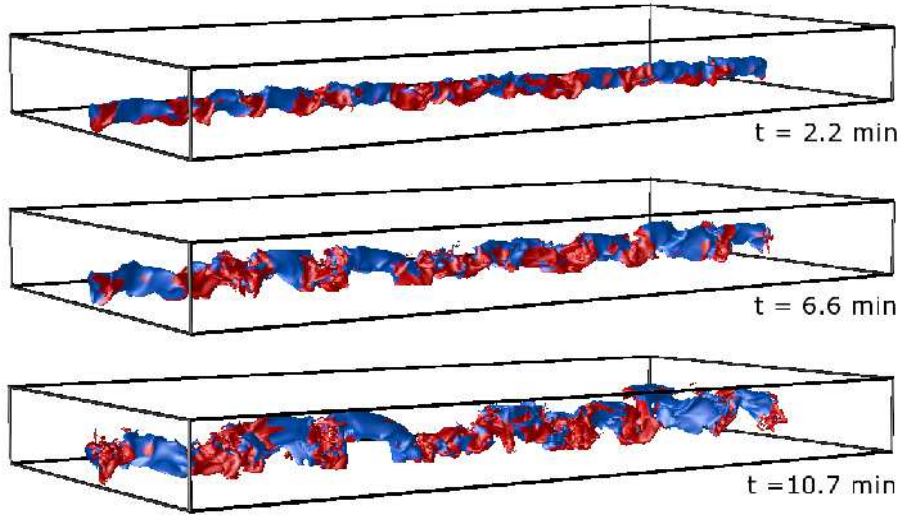}
\caption{The passive evolution of a weak magnetic flux tube (run
U6, $B_0=2500$ G) with the convective flow. Shown is a sequence of
isosurfaces of $|B|=400$ G, on which the blue (red) colour-coding
indicates upflows (downflows). Segments of the tube embedded in
upflows are able to rise and emerge, whereas segments aligned with
downflows are kept submerged.}\label{fig:u5_passive_evolution}
\end{figure*}

Table~1 gives on overview of the various simulation runs carried
out. We start the discussion of the results with the `weak' flux
tube considered in run U6. The initial field strength at the tube
axis is $B_0 = 2500$ G, corresponding to $\beta=8\pi p/B^2\simeq
22$ there. The characteristic radius of the tube is $R_0 = 200$ km
$\simeq 0.4H_p$. The tube is initially twisted, with
$\lambda=0.5$. The longitudinal magnetic flux threading a
cross-section of the tube is $3.1\times10^{18}$ Mx. Criterion
(\ref{ineq:force_balance2}) tells us that the tube must have at
least a field strength of $(H_p/R_0)^{1/2}B_{\rm eq} = 4600-9200$
G in order to rise against the downflows. Since the tube has only
a central field strength of $B_0 = 2500$ G (the average field
strength over its cross-section is even smaller), we expect the
evolution of the flux tube to be dominated by the convective
flows. Fig.~\ref{fig:u5_passive_evolution} shows a time sequence
of 3D isosurfaces of $|B| = 400$ G. In all three panels, the blue
(red) colour-coding indicates that the vertical velocity at the
isosurface is upwards (downwards). Already at $2.2$~min after the
start of the simulation, the flux tube has been significantly
distorted by the convective flows. In the subsequent snapshots,
the upflows have advected segments of the tube upwards, whereas
the downflows have pinned down other segments below the surface.
The shape of the flux tube has developed to a form resembling
a~\emph{sea serpent}. The behaviour of the flux tube in the
presence of asymmetrical up- and down-flows in compressible
convection is reminiscent of the~\emph{turbulent
pumping}~mechanism studied by~\cite{Tobias:TurbulentPumping}.

\begin{figure*}
\centering
\includegraphics[width=0.65\textwidth]{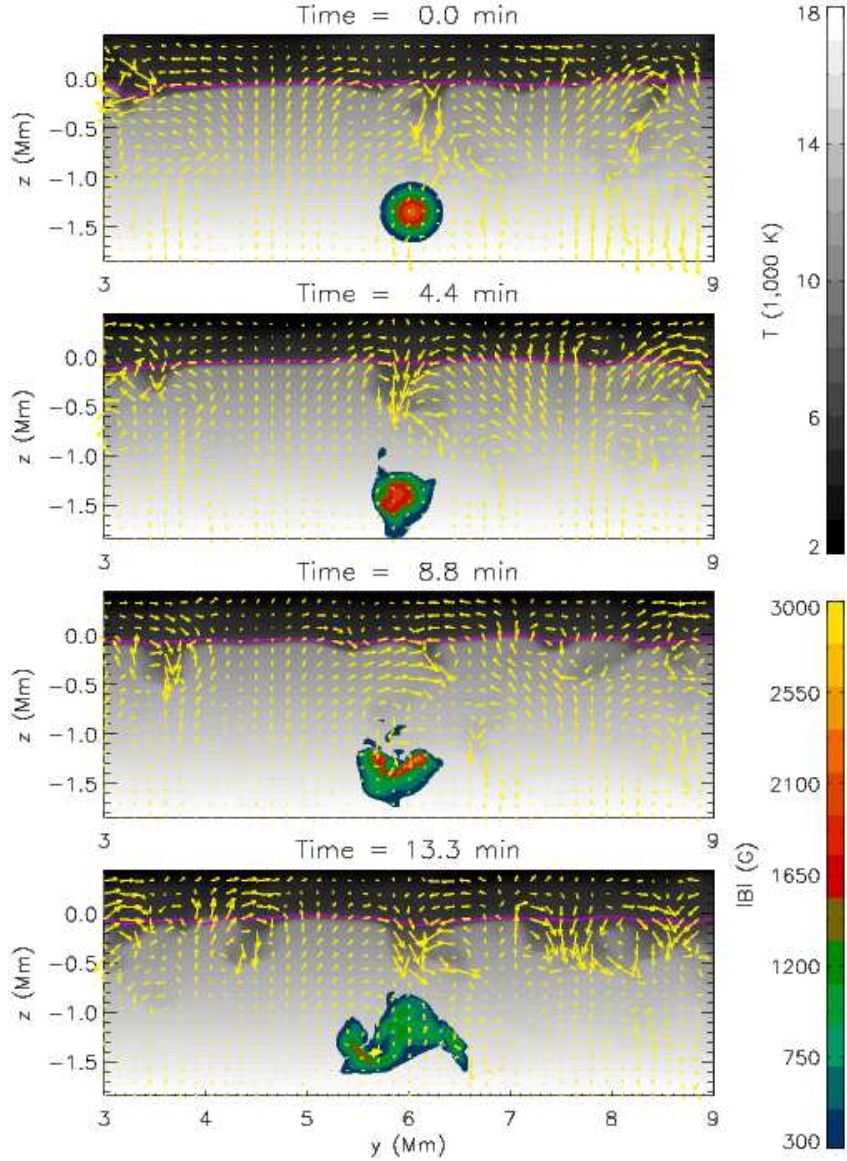}
\caption{Time sequence of a vertical cross-section of the
simulation domain at $x=12$ Mm (perpendicular to the initial tube)
for run U6. The tube segment is passively advected by the
convecting flow. The greyscale indicates the temperature
distribution and the color-coding indicates the distribution of
the absolute field strength, $|B|$. The arrows indicate the
projection of velocity vectors on the vertical cross-section. The
purple line shows the level $\tau_{5000}=1$.}
\label{fig:u5_crosssection}
\end{figure*}

Figure~\ref{fig:u5_crosssection} gives another illustration of how
the flux tube in run U6 evolves passively with the convecting
flow. The figure shows a sequence of snapshots of a vertical
cross-section of the simulation domain at $x=12$ Mm (perpendicular
to the axis of the initial flux tube). The grey scale indicates
the temperature distribution while the colour-coding represents
the absolute field strength, $|B|$. The arrows indicate the
components of the velocity field in the $yz$ plane at $x=12$ Mm.
The purple line running near $z=0$ shows the level of optical
depth unity in the continuum at a wavelength of $5000$
\AA~($\tau_{5000}=1$). At $t=0$ min, this particular segment of
the tube is aligned with a downflow. In the following snapshot at
$t=4.4$ min, we find that this segment of the tube has been
displaced downwards. In the final two snapshots ($t=8.8$ min and
$t=13.3$ min), we witness the tube segment being severely deformed
by the shear in the velocity field at the interface between
upflows and downflows. While the central part of the tube segment
is pushed downwards, other parts of the tube segment are carried
upwards.
\begin{figure*}
\centering
\includegraphics[width=0.8\textwidth]{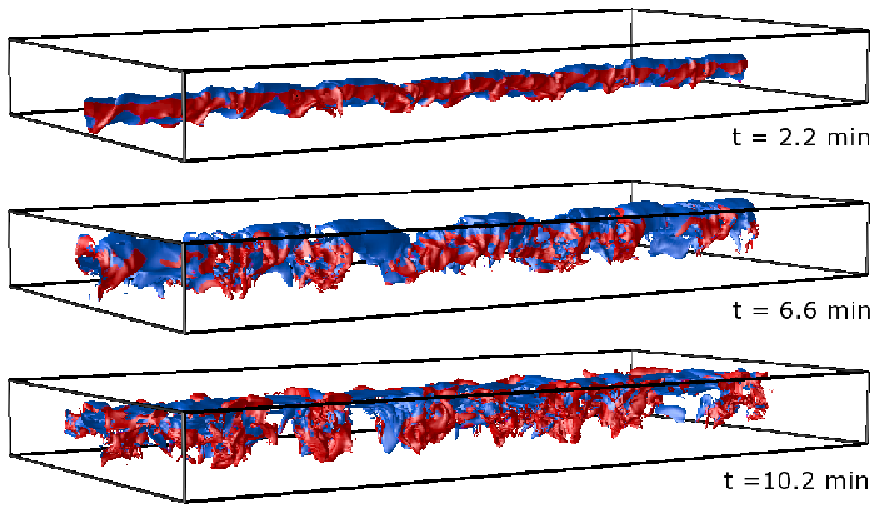}
\caption{Same as Fig.~\ref{fig:u5_passive_evolution} but for a
flux tube with higher initial field strength (run U1, $B_0=8500$
G). The isosurfaces here correspond to $|B|=700$ G. In this case,
the convective flows do not completely control the dynamics of the
flux tube.}\label{fig:u1_active_evolution}
\end{figure*}

\begin{figure*}
\centering
\includegraphics[width=0.65\textwidth]{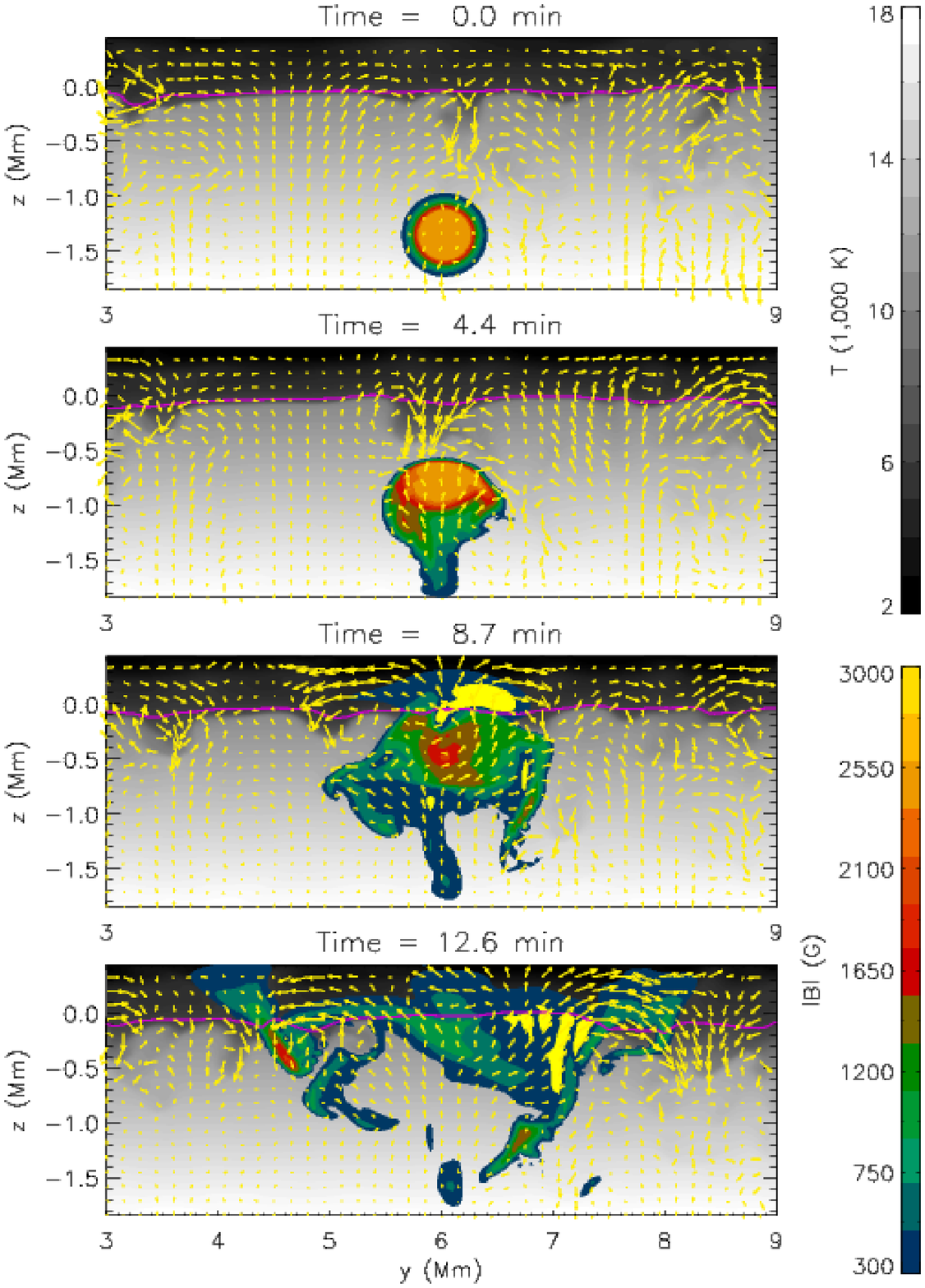}
\caption{Same as Fig.~\ref{fig:u5_crosssection}, but for run U1
($B_0=8500$~G). In this run, the flux tube is sufficiently buoyant
and the magnetic field sufficiently strong to rise against the
drag of the downflow (see snapshot at $t=0$ min). The expansion of
the rising magnetic complex drives strong horizontal flows away
from the emergence site.}\label{fig:u1_crosssection}
\end{figure*}

Our simulation runs with higher initial field strengths confirm
that the ability of the convective flows to dominate the evolution
of the tube lessens with increasing initial field strength. In run
U1, the initial field strength on the axis of the tube is
$B_0=8500$ G, which is within the range of $B_{\rm eq}$ for the
original depth of the tube and corresponds to a $\beta\simeq 1$.
The total longitudinal magnetic flux carried by the tube is
$10^{19}$ Mx. Fig.~\ref{fig:u1_active_evolution} shows a time
sequence of isosurfaces of $|B| = 700$ G for this run. Again, the
colour-coding indicates upwards or downwards vertical velocity. In
this case, the flux tube is not so weak that it is simply advected
by the convective flow. Nor is it sufficiently strong to be
unaffected by the flows. The behaviour of the tube is in an
intermediate regime between these two extremes. On the one hand,
the convective flow is able to deform the tube. On the other hand,
the bulk of the tube finally does overcome the impeding downflows
and emerges at the surface.

Figure~\ref{fig:u1_crosssection} helps emphasize the previous
point. At $t=0$ min, the tube segment shown is aligned with a
downflow. In run U6, this downflow advects the tube downwards. The
situation in run U1 is very different. While the downflow is able
to divert the tube slightly towards the right, it is unable to
keep the tube beneath the surface. Instead, the tube segment is
sufficiently buoyant to overcome the drag by the downflow and
finally emerge.

\subsection{Dependence of emerged flux on initial field strength and twist}
\subsubsection{Unsigned flux at the photosphere}
The `fraction' of the original flux tube which manages to emerge
at the photosphere can be quantified in different ways. One
measure is the maximum value reached in a given time interval by
the total unsigned flux through the plane $z=0$ ($\Phi_{\rm
max}^{z=0}$). For a time interval of $20$~min, this quantity is
indicated in the fifth column of
Table~\ref{table:simulation_runs}. Since a flux tube emerges as an
undulated structure, the emergence actually consists of a
collection of small-scale emergence events, each with a typical
length-scale of about $1$ Mm. This means that the unsigned
vertical flux at the photosphere can be several times that of the
initial longitudinal flux of the tube ($\Phi_0$). We identify two
clear trends. Firstly, $\Phi_{\rm max}^{z=0}$ increases with
increasing $B_0$ for constant $\lambda$. This simply reflects the
fact that a magnetic flux tube with higher $B_0$ contains more
flux (for the same tube radius) and is more buoyant. Secondly,
$\Phi_{\rm max}^{z=0}$ increases with $\lambda$. This is due to
the fact that the component of the magnetic field transverse to
the original tube axis contributes to the unsigned flux at $z=0$.

\subsubsection{Emergence of tracer fluid elements}
\begin{figure*}
\centering \subfigure[Dependence on
$B_0$]{\includegraphics[width=0.4\textwidth]{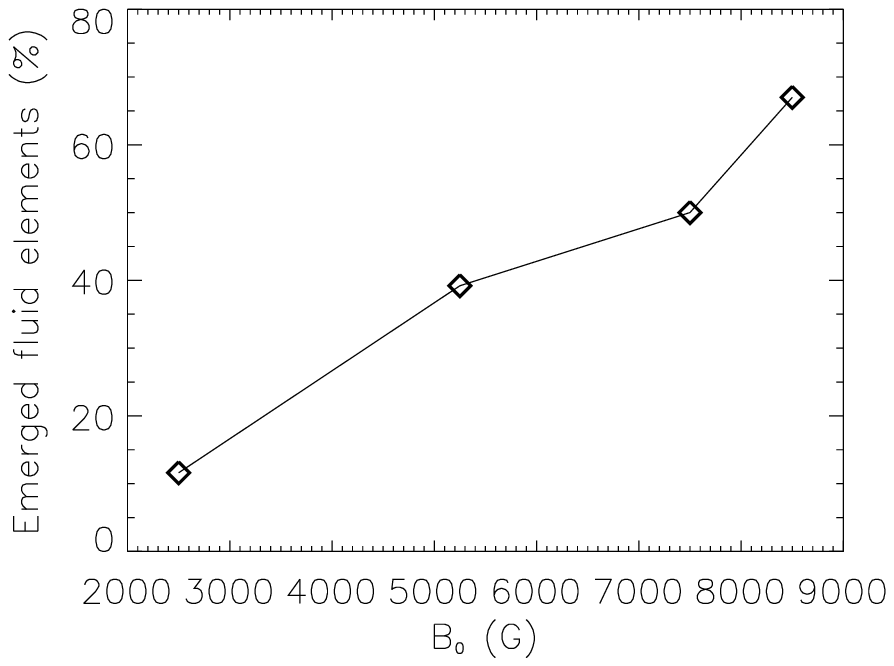}}
\subfigure[Dependence on twist parameter
$\lambda$]{\includegraphics[width=0.4\textwidth]{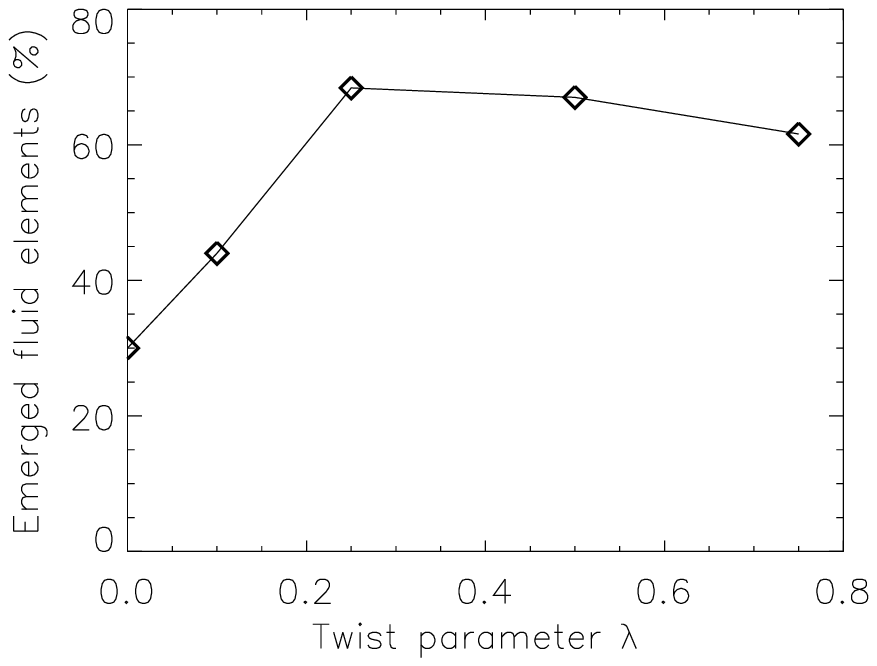}}
\caption{Percentage of Lagrangian fluid elements initially
embedded in the magnetic flux tube that reach the base of the
photosphere ($z=0$) within the time interval $0 \le t \le 20$
min.}\label{fig:lagangian_emergence}
\end{figure*}

Another measure of the fraction of the tube that emerges can be
obtained by considering selected (Lagrangian) fluid elements as
tracers of the evolution. To this end, we followed the
trajectories of $250$ such tracer fluid elements (referred to
simply as `tracers' in what follows). The first $50$ of them were
located equidistantly along the axis of the initial flux tube. The
remaining tracers were selected similarly to lie along four lines
parallel to the tube axis, each line at a distance of $50$ km from
the axis (one line lying above the axis, one lying below, and the
other two lines to the left and right of the axis). We determined
the fraction of tracers that manage to reach the photosphere
($z\ge 0$) within the time period $0 \le t \le 20$ min. The
numerical values of the results indicated in the last column of
Table~\ref{table:simulation_runs} and plotted in
Fig~\ref{fig:lagangian_emergence}. For flux tubes with the same
initial twist $\lambda$, the percentage systematically increases
with the initial field strength, $B_0$. As a reference case, we
performed a non-magnetic run (i.e. $B_0 = 0$) and found that $7\%$
of the fluid elements emerged within $20$ minutes. In the case of
a weak magnetic flux tube (run U6, $B_0=2500$ G), $12 \%$ of the
tracers emerged, while for a strong tube (run U1, $B_0 = 8500$ G),
$67\%$ of the
 tracers managed to emerge. This trend is due to the fact that the
buoyancy of a flux tube scales with $B^2$.

The dependence of the percentage of emerged tracers on the twist
parameter $\lambda$ is not so straightforward. We see that, at a
given magnetic field strength of $B_0=8500$ G, the percentage
increases from $30\%$ to $68\%$ in the range $\lambda=0-0.25$.
This trend is compatible with results from earlier idealized
simulations of rising magnetic tubes. These studies found that an
untwisted tube, after rising a distance of a few times its
diameter, will split into counter-rotating vortex rolls which
separate horizontally from each
other~\citep*{Schuessler:BuoyantMagFluxTubes,LongcopeFisherArendt}.
The fragments of the individual tubes have smaller radii, less
flux and are less buoyant. As a result, they are more easily
disrupted by the external flow. Simulations of twisted magnetic
tubes show that the magnetic tension of the transverse field tends
to keep the tube coherent
\citep*{Moreno-InsertisEmonet:RiseOfTwistedTubes, Fan:2DTubes,
EmonetMoreno-Insertis:PhysicsOfTwistedFluxTubes,
Cheung:MovingMagneticFluxTubes}. This is why the percentage of
emerged tracers increases from $\lambda=0$ to $\lambda=0.25$. For
$\lambda>0.25$, however, the percentage drops again. This reversal
in the trend is not in contradiction with the previous
explanation. The complicating factor here is the relationship
between the twist of a tube and its buoyancy. For the initial
conditions we have chosen (i.e. the divergence of the total stress
tensor at $t=0$ closely resembling that of the non-magnetic
domain), the density deficit at the core of the tube scales as
$\Delta \varrho/\langle \varrho \rangle\sim 1/(1+\lambda^2/2)$.
Consequently, for a given initial field strength, the density
deficit in the core of the tube actually {\em decreases} with
increasing $\lambda$. Hence, although the flux tube remains more
coherent with increasing $\lambda$, it is also somewhat less
buoyant than the untwisted case.

\subsection{Trajectories of tracer fluid elements}
\label{subsec:tracerelements}

\begin{figure*}
\centering {\includegraphics[width=0.4\textwidth]{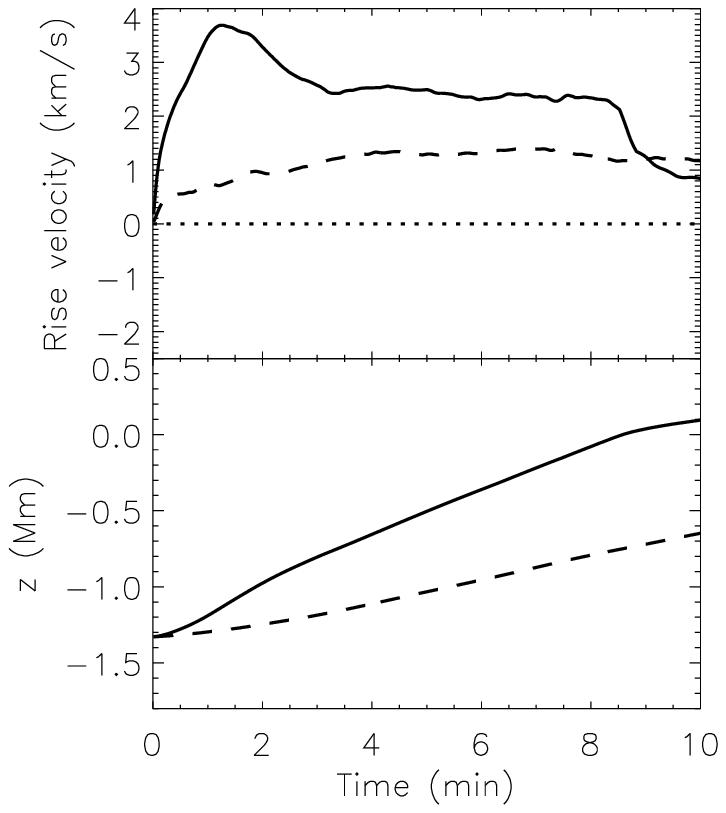}}
\centering {\includegraphics[width=0.4\textwidth]{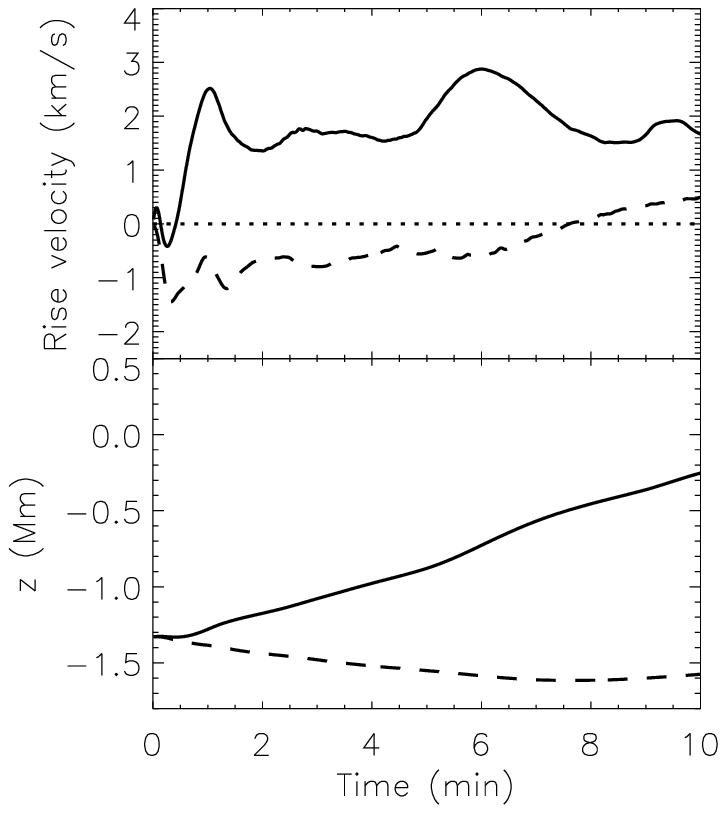}}
\caption{Motion of tracer fluid elements initially located at the
axis of the magnetic flux tube. The left panel shows time profiles
of rise velocity and height of a fluid element in a segment of the
tube which was initially aligned with a convective upflow. The
right panel shows the corresponding profiles for a fluid element
in a tube segment in a convective downdraft. The solid and dashed
lines indicate profiles for runs U1 and U6, respectively.}
\label{fig:t1t5comparison}
\end{figure*}

Figure~\ref{fig:t1t5comparison} shows the vertical velocity and
height of two tracers initially located at the axis of the
magnetic flux tube. Solid and dashed lines correspond to results
from run U1 and U6, respectively. The left panel corresponds to a
tracer representing a segment of the initial flux tube that was
embedded in a convective upflow. While the tracers in both runs
accelerate and rise towards the surface, the tracer in run U1
(solid line) experiences larger acceleration and reaches higher
rise speeds ($v_y \approx 2-3$ \kms) than the tracer in run U6
($v_y\approx 1$ \kms). This difference can be attributed to the
higher buoyancy of the flux tube in run U1. The right panel of
Fig.~\ref{fig:t1t5comparison} shows the corresponding results for
a tracer in a tube segment initially located in a downdraft. Here,
the tracer in run U6 is forced by the downdraft to descend by
about 300 km before it gets caught up in a neighbouring upflow.
The tracer in run U1 is also initially pushed downwards by the
convective downdraft, but the flux tube in this case is
sufficiently buoyant to counteract the downflow.

\begin{figure}
\centering
\includegraphics[width=0.4\textwidth]{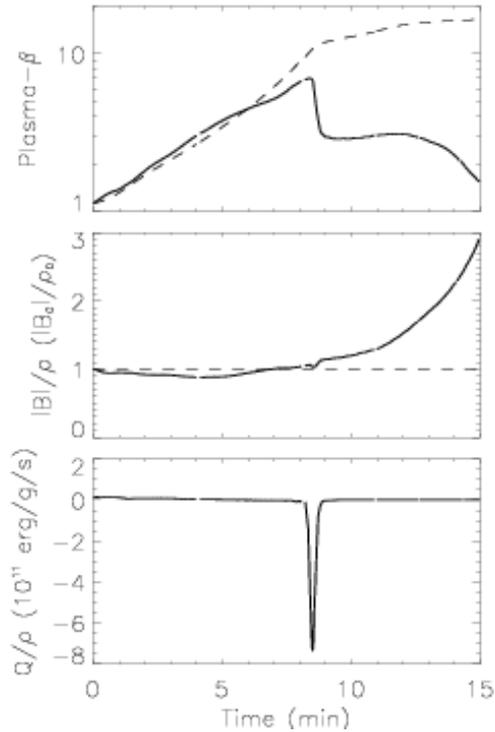}
\caption{Temporal evolution of plasma $\beta$, ratio
$|B|/\varrho$, and specific radiative heating, $Q_{\rm
rad}/\varrho$, for a tracer fluid element initially located at the
axis of the tube segment embedded on a convective upflow (solid
curves). The dashed curves indicates values corresponding to a
thin flux tube model assuming adiabatic motion and no variation
along the tube axis. As long as there is negligible radiative
heating and draining of mass at the position of the tracer, there
is a good match between the thin-tube model and the actual
profiles.} \label{fig:thin_tube_comparison}
\end{figure}

Fig.~\ref{fig:thin_tube_comparison} shows the time evolution of
some further properties of the tracer initially embedded in an
upflow (Fig.~\ref{fig:t1t5comparison}, left panel): plasma $\beta$
(top panel), $|B|/\varrho$ (middle), and $Q_{\rm rad}/\varrho$,
the radiative heating rate per unit mass. Until $t\simeq8$~min,
the tracer fluid element rises almost adiabatically ($Q_{\rm
rad}\approx 0$). During this time, as the fluid rises and expands,
the value of $\beta$ increases from an initial value of $1$ to a
maximum value of $7$. Thereafter, at about $t=8.5$ min, the fluid
element reaches the transparent photosphere and radiates
intensely. The corresponding loss of internal energy corresponds
to a decrease of the local gas pressure, leading to a compression
of the fluid element and an abrupt decrease of $\beta$ to a value
of $3$. Thereafter, the Lagrangian tracer continues to rise
through the photosphere (eventually escaping through the top
boundary of the domain). As it enters the more tenuous layers of
the photosphere, it expands. Despite this, the plasma-$\beta$
hovers for a few minutes at the level of $\beta\approx 3$ and then
decreases even further to a value approaching unity. The time
profile of the ratio $|B|/\varrho$ (middle panel of
Fig.~\ref{fig:thin_tube_comparison}) provides a hint as to why
this happens. We see that after the fluid has entered the
photosphere, $|B|/\varrho$ begins to increase. By virtue of
Walen's Equation (derived by combining the equation of continuity
and the ideal induction equation, thus ignoring magnetic
diffusion),
\begin{equation}
\frac{D}{Dt}\left( \frac{\vec{B}}{\varrho}\right) =
\frac{1}{\varrho}(\vec{B} \cdot \nabla) \vec{v},
\end{equation}
we see that $|B|/\varrho$ increases in the case of stretching of
fluid along field lines and decreases in the case of
de-stretching.

We can compare the evolution of the tracer (which was initially
embedded at the tube axis) with what would result from the simple
model of a horizontal thin flux tube rising adiabatically in a
stratified, static atmosphere. The result is shown by the dashed
curves in Fig.~\ref{fig:thin_tube_comparison}. The assumption that
there are no variations along the direction of the tube axis means
that $|B|/\varrho$ is constant (flat dashed line in the middle
panel). Assuming adiabatic rise and instantaneous total pressure
balance with the (prescribed) surroundings of the thin tube, we
can determine the evolution of the properties of the thin flux
tube as functions of its initial state and its vertical
displacement \citep{Cheung:MovingMagneticFluxTubes}. Between $t=0$
and $t=8$ min, there is very good agreement between the actual and
thin-tube model profiles, resulting from the quasi-adiabatic
motion and negligible amount of mass drainage or accumulation
along the field lines. Once the Lagrangian tracer reaches the
surface and loses entropy, the profiles of $\beta$ diverge.
Whereas the actual value of $\beta$ decreases sharply (due to
cooling) and then more gradually (due to mass drainage), the
thin-tube model predicts that $\beta$ continues to rise smoothly
as the fluid continues to rise in the photosphere. The actual and
model profiles of $|B|/\varrho$ also diverge. Since our thin-tube
model does not take into account variations along the tube,
$|B|/\varrho$ remains constant in the model profile. In the actual
profile, $|B|/\varrho$ increases after the fluid elements has
emerged. This is ultimately a result of radiative cooling, which
makes the emerged fluid denser than the fluid immediately below,
driving the development of a downflow. This accelerating material
in the developing downflow stretches the fluid along the field
line and increases $|B|/\varrho$.

\section{Observational signatures}
\label{sec:observational_signatures} In the following sections, we
compare diagnostics derived from the simulation results with
observations of flux emergence.

\subsection{`Quiescent' flux emergence}
\label{subsec:quiescent_flux_emergence}

The tube in run U6 evolves passively with the convective flow.
Consequently, the emergence of magnetic flux does not lead to a
severe disturbance in the appearance of the granulation in white
light intensity images. For this reason, we refer to these flux
emergence events as~\emph{quiescent}.

\begin{figure*}
\centering
\includegraphics[width=0.75\textwidth]{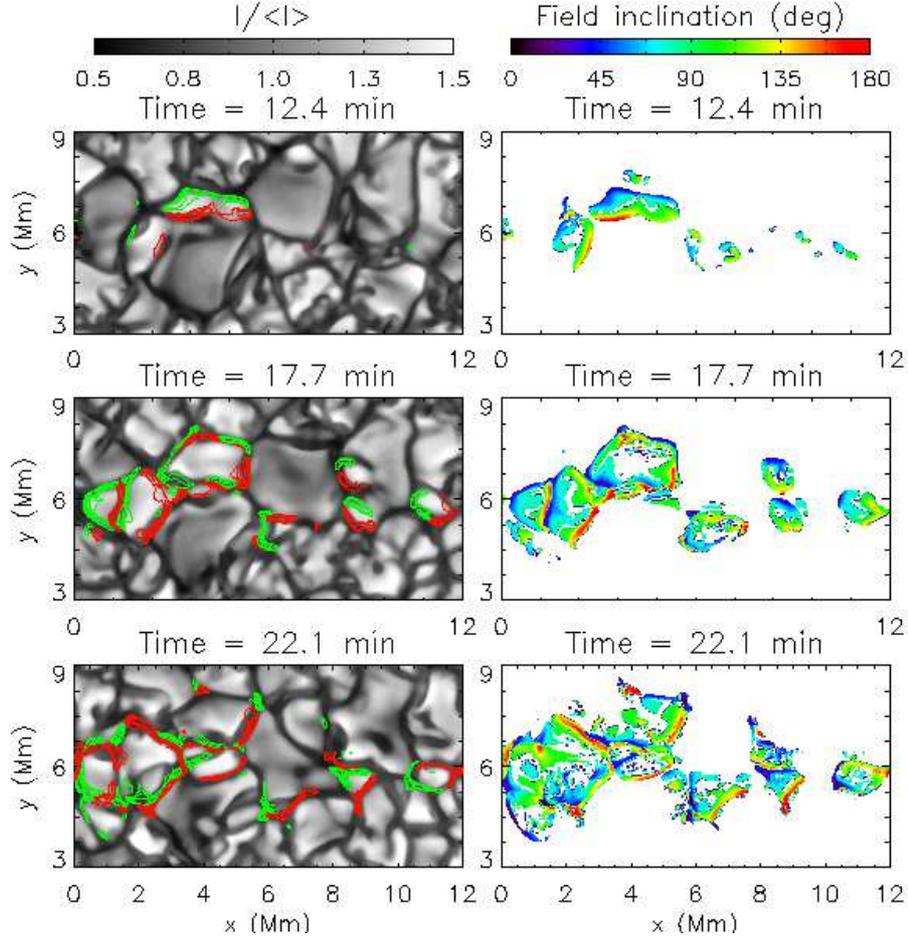}
\caption{Emergence of magnetic field into the photosphere in run
U6.~\emph{Left column}:  time sequence of the vertically emerging
continuum intensity at 5000~\AA, $I_{\rm 5000}$.~\emph{Right
column}: the corresponding time sequence of magnetic field
inclination evaluated at $\tau_{\rm 5000}=0.1$. A field
inclination (also called zenith angle) of $0^\circ$ and
$180^\circ$ respectively correspond to vertical magnetic field
directed out of and into the $x-y$ plane. In the left panels, the
color contours indicate the vertical component of the magnetic
field at $\tau_{\rm 5000}=0.1$ with the levels $\pm
[50,100,200,400]$~G. Green (red) contours indicate magnetic upward
(downward) directed field. An accompanying mpeg animation is
available at http://www.mps.mpg.de/homes/cheung/U6emergence.mpg.}
\label{fig:u5_magnetograms}
\end{figure*}

Figure~\ref{fig:u5_magnetograms} shows a sequence of continuum
intensity and magnetic field maps representing the flux emergence
event in run U6. The magnetic field is evaluated at
$\tau_{5000}=0.1$ (optical depth of $0.1$ for the visible
continuum at $5000$ \AA). We chose to sample from this optical
depth because the maxima of the contribution functions of many
photospheric lines commonly used to probe surface magnetic fields
(e.g.~Fe~I~$6301$, Fe~I~$6302$) are roughly located at this depth.
As such, we can expect that the information carried by the line to
be representative of the physical quantities in this layer of the
atmosphere. In the left panels of Fig.~\ref{fig:u5_magnetograms},
the overlaid colored contours indicate the vertical component of
the magnetic field ($B_z$).

The flux tube is initially aligned in the $x$-direction at $y=6$
Mm. At $t=12.4$ min, we see the first signs of flux emergence at
the surface. The magnetic field emerges within the interior of
granules, being predominantly horizontal near the granular centre.
Towards the edge of the granule, the magnetic field becomes more
vertical, indicating that the magnetic field threading a granule
is forced to emerge in an arched configuration. One example is the
bipolar region emerging in the granule centered at $[x,y]=[4,7]$
Mm (see snapshot at $t=12.4$ min). At this particular instant, the
flux contained in each polarity of this small bipole is $8\times
10^{17}$ Mx. The diverging horizontal flow of the granule flow has
the effect of expelling the emerged flux into the network of
intergranular downflow lanes within a typical granulation
timescale on the order of $5$~min
\citep{BrayLoughhead:SolarGranulation,Title:SolarGranulation}.
This finding is in accordance with earlier simulations of
magneto-convection in an idealized
setup~\citep{Weiss:FluxExpulsion, Proctor:Magnetoconvection1982,
Hurlburt:Magnetoconvection}~as well as more recent, realistic
simulations of magneto-convection in the near-surface layers of
the solar convection zone~\citep{Steiner:etal:1998,Bercik:PhD,
Bercik:Magnetoconvection, Voegler:MURaM,
Stein:SmallscaleMagnetoconvection}. At $t=17.7$ min (about $5$
minutes after the first snapshot), the distribution of the
vertical field $B_z$ already shows that the regions of strongest
field are in the intergranular lanes. However, the surface field
strength at $\tau_{\rm 5000}=0.1$ rarely exceeds the equipartition
value of $450$ G. In comparison, the granule interiors have
significantly weaker fields. The next snapshot at $t=22.1$ min
illustrates the further transport and dispersion of the emerged
flux in the intergranular network.

\begin{figure*}
\centering
\includegraphics[width=0.7\textwidth]{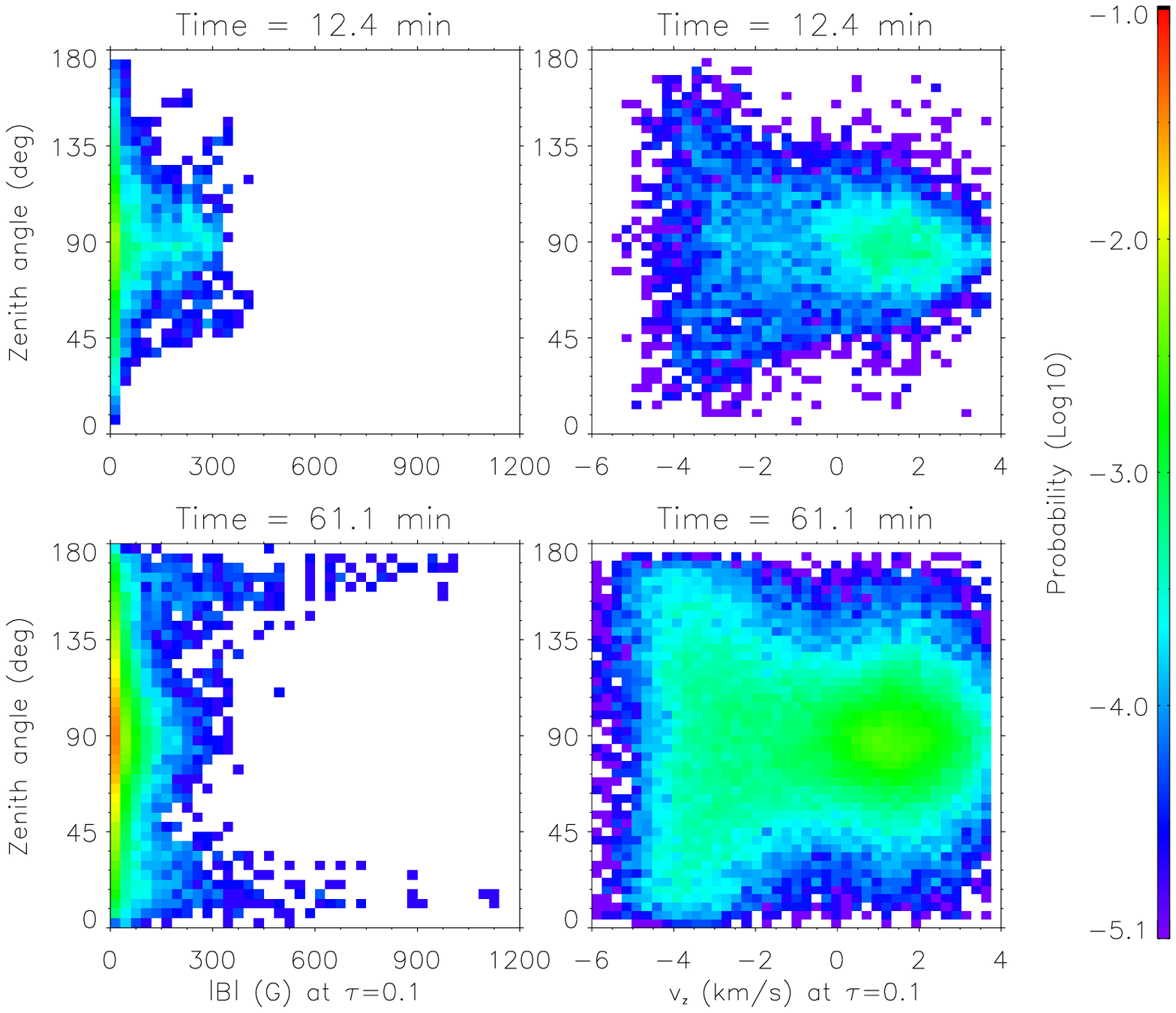}
\caption{\emph{Left column}: Joint probability distribution
functions (JPDFs) of the zenith angle of the magnetic field vector
and the absolute field strength, $|B|$, both at the level
$\tau_{\rm 5000}=0.1$. \emph{Right column}: JPDF of zenith angle
and vertical fluid velocity. The colour coding indicates the
logarithm of the probability. The upper and lower rows,
respectively, show the JPDFs at $t=12.4$ min and $t=61.1$ min for
run U6.}\label{fig:u5_inclination}
\end{figure*}
How do our simulation results compare with observations?
\citet{DePontieu:Small-scaleEmergingFlux} has obtained high
cadence and high resolution observations of an emerging flux
region (EFR). Within a time series of $3.5$~hr duration, he
identified seven individual small-scale emergence events. His two
key findings relevant for this discussion are: {(a)}~magnetic
concentrations emerge in the interior of granules; and
{(b)}~within $10-15$ min from their initial appearance, the flux
concentrations are quickly dispersed by the granular flow. He
estimates the average flux density of the emerging magnetic
concentrations to be about $200\pm30$ Mx cm$^{-1}$. If this value
were to correspond to the actual magnetic field strength, the
emerging fields would have sub-equipartition strength ($B_{\rm eq}
\approx 450$ G), a finding consistent with his results {(a)} and
{(b)}. ~\citeauthor{DePontieu:Small-scaleEmergingFlux} also gives
estimates for the average flux of the emerging concentrations in
the range $(9 \pm 4) \times10^{17}$ Mx. This is the same order of
magnitude as what we find for individual polarities of emerging
flux concentrations in run U6.

\citeauthor{DePontieu:Small-scaleEmergingFlux} suggested that the
small-scale emergence events he has detected may be related to the
so-called \emph{horizontal internetwork
fields}~\citep[HIFs,][]{Lites:HIFs}, which are predominantly
horizontal magnetic structures with angular and time scales of
$1-2''$ and $5$~min, respectively. The existence of HIFs are
inferred from the distinctive signals in the Stokes Q and U
profiles of relatively quiet regions of the solar surface. The
strength of these signals indicate that HIFs are generally weak
with $|B| \lesssim 600$ G. HIFs are usually detected in regions
where the spectral lines are slightly blueshifted, indicating an
association with upflows. ~Both~\citeauthor{Lites:HIFs}
and~\citeauthor{DePontieu:Small-scaleEmergingFlux} interpret these
as signatures of the crests of magnetic loops emerging in
granules.

We have analyzed simulation run U6 to check whether this
interpretation is consistent with our results.
Fig.~\ref{fig:u5_inclination} shows joint probability distribution
functions (JPDFs) of different surface quantities in the
simulation run for times $t=12.4$ min (upper row) and $t=61.1$ min
(lower row). The left column shows JPDFs between the~\emph{zenith
angle}~$\gamma$ of the magnetic field and the absolute field
strength, $|B|$, both evaluated at $\tau_{5000}=0.1$. The zenith
angle is defined as the angle between $\vec{B}$ and the vertical
direction ($z$-axis). As such, magnetic field vectors with
$\gamma=0^\circ$ and $\gamma=180^\circ$ correspond to purely
vertical magnetic field that point out of and into the horizontal
plane, respectively. Purely horizontal fields have
$\gamma=90^\circ$. The right column shows JPDFs of the zenith
angle of $\vec B$ and the vertical fluid velocity (for surface
regions with $|B|>0$). Flux begins to emerge approximately at
$t=12.4$ min. The JPDFs at this time indicate the existence of
predominantly horizontal field with strengths of up to $|B|
\approx 400$ G. These horizontal fields mainly have rise
velocities of $1-2$~\kms, which is typical of granular upflows.
This finding is compatible with the scenario proposed
by~\citet{Lites:HIFs} and
by~\citet{DePontieu:Small-scaleEmergingFlux} for the explanation
of HIFs.

The snapshot taken at $t=61.1$~min (lower panels in
Fig.~\ref{fig:u5_inclination} shows a distinctly different shape
for the JPDF between the zenith angle and $|B|$. The pair of
horn-like features indicate strong vertical fields of up to $1$
kG. We will return to this point in the following section.
Furthermore, there is a continuous distribution of nearly
horizontal fields with strengths of up to $300$~G. The
accompanying JPDF between the zenith angle and $v_z$ indicates
that the majority of horizontal fields reside within upflow
regions (granules). In fact, the JPDF between the zenith angle and
$|B|$ at $t=61.1$ min resembles those found in simulations of
magneto-convection in the near-surface layers with a net vertical
magnetic flux threading the simulation domain \citep{Voegler:PhD}.
This indicates that the abundance of weak horizontal fields in
granule interiors is a general feature of magneto-convection and
not necessarily always associated with emerging flux.

\subsubsection{Surface evolution of emerged field: cancellation,
               coalescence and secondary emergence}
\label{subsubsec:surface_evolution} In run U6, the morphology of
the emerged field bears little resemblance to the initial
horizontal flux tube structure. The arrangement of the surface
field after the initial flux emergence (see snapshots at $t=17.7$
and $22.1$ min in Fig.~\ref{fig:u5_magnetograms}) resembles the
`salt and pepper' pattern of quiet Sun magnetic fields observed,
e.g., by \citet{DominguezCerdena:InternetworkMagneticFields2003},
displaying a mixture of positive and negative small-scale flux in
the intergranular network. These authors describe instances when
flux concentrations of like polarities meet and appear to coalesce
(to within the resolution limit). They also show a flux
cancellation event due to the encounter of opposite polarities. In
our simulation run U6, both these types of events occur as well.
In the following, we discuss in detail an instance of a
coalescence event between vertical flux concentrations of
like-polarity.

\begin{figure*}[t]
\centering
\includegraphics[width=0.65\textwidth]{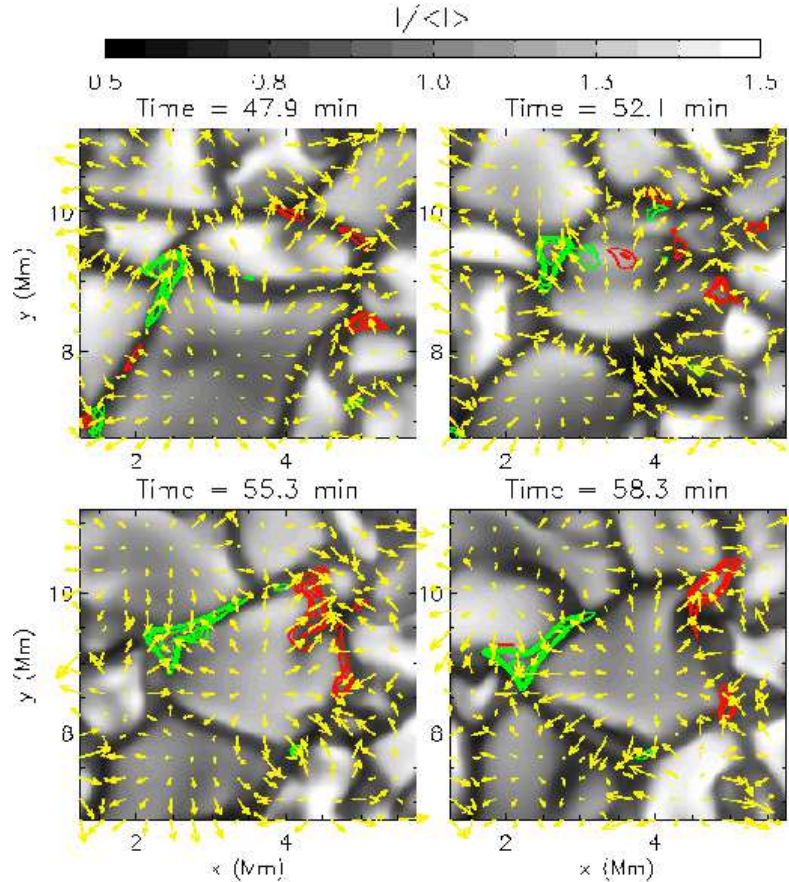}
\caption{Coalescence of newly emerged magnetic concentrations with
pre-existing surface field in simulation run U6. The grey shading
in each panel represents the normalized continuum intensity at
$5000$~\AA. The vector overlays indicate the horizontal velocity
field at optical depth $\tau_{\rm 5000}=0.1$ and the contour
overlays indicate $B_{z}$ at the same optical depth at the levels
$\pm [100,200,400,800,1600]$ G. Green (red) contours correspond to
positive (negative) polarities. At $t=52.1$~min, a bipole centered
at $[x,y]=[3.5,9.2]$ Mm is emerging through a granule. Flux
expulsion leads to coalescence of the pre-existing positive flux
with the positive part of the newly emerged
flux.}\label{fig:u5_coalescence_event}
\end{figure*}

Figure~\ref{fig:u5_coalescence_event} shows a time sequence of
intensity images in the vicinity of a granule, overplotted with
horizontal velocity vectors and contours of the vertical magnetic
field strength, both at $\tau_{\rm 5000}=0.1$. At time $t=47.9$
min, we find a number of magnetic flux concentrations of both
polarities. The most prominent feature is centered at
$[x,y]=[2.5,9]$ Mm. This feature resides at a downflow vertex of
intergranular lanes and has a flux of $1.5 \times 10^{18}$ Mx. The
core of the feature has a maximum vertical field strength of
$B_z=1,200$ G (at $\tau_{\rm 5000}=0.1$). A few minutes later, at
$t=52.1$ min, we find a small bipole emerging through a granule
that is located to the right of the pre-existing flux
concentration. The bipole is oriented such that the pole closest
to the pre-existing flux concentration has the same polarity
(positive). The newly emerged flux is subsequently expelled to the
adjacent downflow lane, where it coalesces with the pre-existing
feature. The resulting larger flux concentration has a net flux of
$2.6 \times 10^{18}$ Mx, comparable to the longitudinal flux of
the horizontal flux tube at the beginning of the simulation
($\Phi_0=3.1\times10^{18}$ Mx).

This particular example is interesting for a number of reasons.
Firstly, the new bipole emerges at a relatively late stage of the
simulation, about $40$ min after the initial appearance of flux at
the surface (see Fig.~\ref{fig:u5_magnetograms}). Its emergence
location is also unusual. In Fig.~\ref{fig:u5_magnetograms}, we
see that almost all the flux emerges between $y=4$ and $y=8$ Mm.
The new bipole, however, emerges significantly further away, at $y
= 9.2$ Mm. These two points are consistent with the fact that,
when a section of a flux tube is passively carried by an
upwelling, not all the flux contained in the tube emerges in one
single event. The material near the edge of an upwelling may
overturn before it reaches the surface. Consequently, the magnetic
field threading this material fails to emerge on first attempt.
This magnetic field may then continue to travel downwards, or may
get caught up in another upwelling and travel upwards again. The
emergence of the bipole in Fig.~\ref{fig:u5_coalescence_event} is
one such example. The magnetic field associated with this bipole
has overturned several times before it eventually reaches the
$\tau_{5000}=0.1$ surface. Considering that it had been caught up
in different upflows and downflows, it is not surprising that its
emergence location is about $1$ Mm further afield.

\begin{figure*}
\centering
\includegraphics[width=0.8\textwidth]{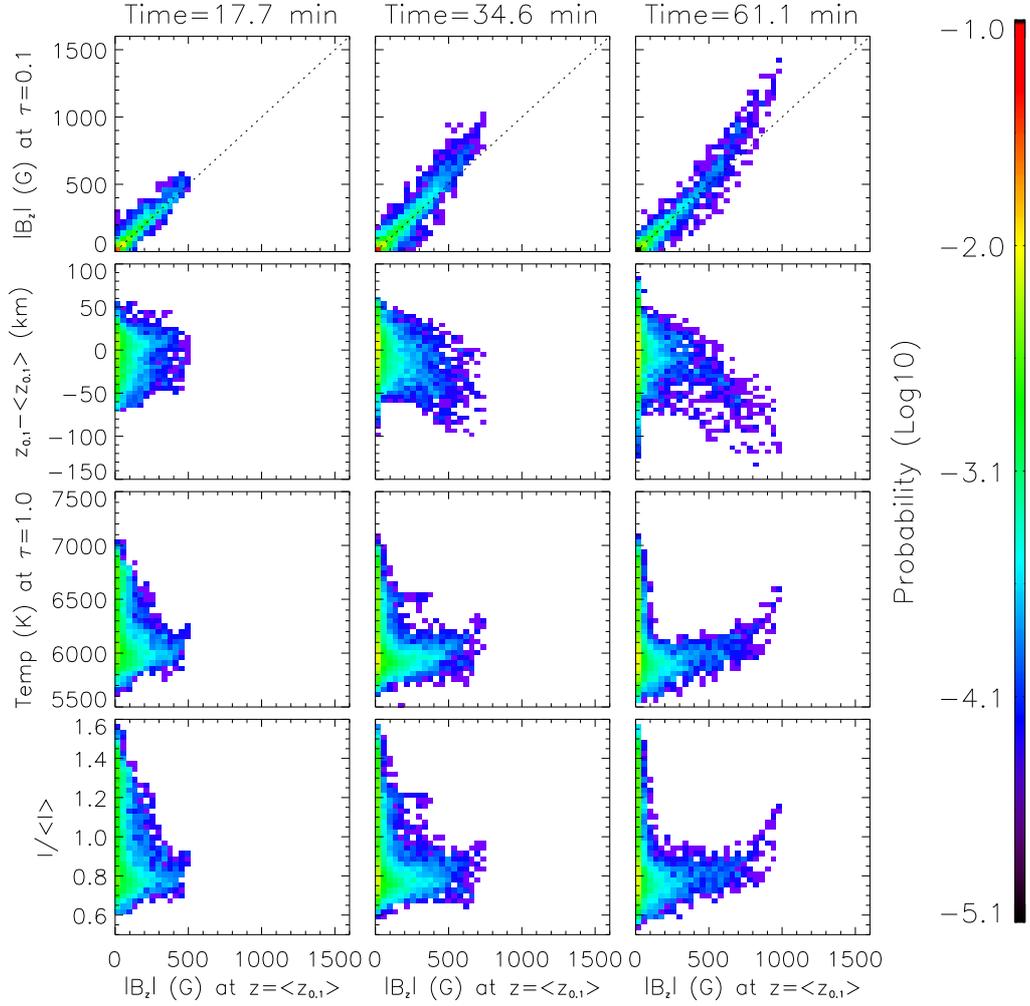}
\caption{Joint probability distribution functions (JPDFs) for
various quantities in run U6. In all panels, the horizontal axis
gives the value of $|B_z|$ at a plane of constant geometrical
height, $z=\langle z_{0.1}\rangle$. The quantities given by the
vertical axis are: $|B_z|$ at the local level $\tau_{5000}=0.1$
(first row of panels), the local vertical displacement of the
$\tau_{5000}=0.1$ surface from the mean geometrical height of the
surface corresponding to this optical depth ($z_{0.1}-\langle
z_{0.1} \rangle$, second row), the temperature at
$\tau_{5000}=1.0$ (third row), and the normalized emergent
continuum intensity at $5000$~\AA (fourth row). The three columns
show the JPDFs at three different times. The color coding
indicates the logarithm of the probability.}
\label{fig:u5_histograms}
\end{figure*}

The coalescence event shown in Fig.~\ref{fig:u5_coalescence_event}
is interesting for another reason. The pre-existing positive flux
concentration residing in a downflow vertex is not as dark as the
average intergranular network. After coalescence, the resulting
concentration becomes even brighter. To explain this, we make
reference to Fig.~\ref{fig:u5_histograms}. Let us first consider
the top row of this figure, which shows joint probability
distribution functions (JPDFs) between the value of $|B_z|$ at a
horizontal plane of constant geometrical height $\langle
z_{0.1}\rangle$ and its value evaluated at optical depth
$\tau_{5000}=0.1$. The height $\langle z_{0.1} \rangle$ is defined
as the average geometric height of the $\tau_{5000}=0.1$ surface
in the absence of magnetic fields. On average, the
$\tau_{5000}=0.1$ surface lies about 160~km above the level
$\tau_{5000}=1.0$. We first focus on the distribution of field
strengths at the horizontal plane $z=\langle z_{0.1}\rangle$ (i.e.
ignore the y-axis). A few minutes after the initial appearance of
flux at the surface ($t=17.7$ min), the vertical field strengths
do not exceed $500$ G. Some time later, at $t=34.6$ min, a small
fraction of the vertical flux concentrations have reached field
strengths of up to $700$ G, and at $t=61.1$ min we find fields
approaching $|B_z| = 1$ kG at $z=\langle z_{0.1}\rangle$.

The appearance of stronger magnetic fields several granulation
timescales after the initial emergence of flux suggests that
convective intensification operates
\citep[e.g.,][]{Nordlund:Stein:1990, Grossmann-Doerth:etal:1998,
Voegler:MURaM, Stein:SmallscaleMagnetoconvection}. The magnetic
field maps from run U6 show that the maximum value of $|B_z|$
within a vertical magnetic bundle is an increasing function of its
total magnetic flux, $\Phi$. In order to identify individual flux
bundles, we choose a threshold value, $B_{\rm th}$. For each
discrete region in the magnetic field map with $|B_z| \geq B_{\rm
th}$, we measured the maximum value of $|B_z|$ in the region as
well as the amount of flux contained within the region.
\begin{figure}
\centering
\includegraphics[width=0.4\textwidth]{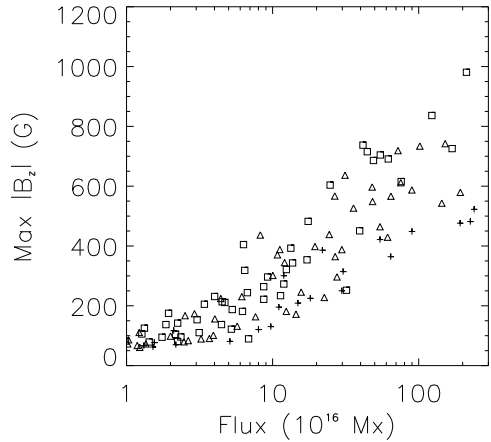}
\caption{Scatter plot of the maximum value of $|B_z|$ in
individual magnetic flux concentrations versus the amount of
vertical flux contained in the flux concentration. The values of
$|B_z|$ were evaluated at a plane of constant geometrical height,
$z=\langle z_{0.1}\rangle$. The crosses, diamonds and squares
indicate values corresponding to the snapshots taken at
$t=17.7$~min, $t=34.6$~min, and $t=61.1$ min, respectively.}
\label{fig:u5_convective_intensification}
\end{figure}
A scatter plot of these two quantities is shown in
Fig.~\ref{fig:u5_convective_intensification}. The crosses,
diamonds and squares correspond to the snapshots taken at
$t=17.7$~min, $t=34.6$~min, and $t=61.1$~min, respectively, for
$B_{\rm th}=50$ G. The scatter plot clearly shows that larger flux
bundles can support stronger internal field strengths. This trend
is not sensitive to the threshold value used. Our finding is
compatible with the work
of~\citet{Venkatakrishnan:InhibitionOfConvectiveCollapse} - who
suggested that the radiative heating of flux tubes by their
surroundings inhibits the intensification of very small flux tubes
- and with the observational study by
\citet{Solanki:IR_12_CC_inhibition}. We should point out, however,
that the various types of diffusion present in the simulation
affect the results for the smallest flux concentrations. For
instance, a flux bundle with a mean field strength of $100$ G and
a flux of $10^{16}$ Mx has a radius of about $50$ km, equal to the
horizontal grid spacing used in our simulations, so that it is
essentially unresolved. Values in the scatter plot in the range
$\Phi \lesssim 10^{17}$~Mx are therefore probably influenced by
diffusive effects.

For the snapshot taken at $t=61.1$~min, the maximum field strength
at $z=\langle z_{0.1}\rangle$ has reached about 1~kG. The
corresponding magnetic pressure is almost equal to the
horizontally averaged gas pressure at that geometrical level. This
indicates that the strongest flux concentrations are largely
evacuated, leading to a downward shift of the optical depth scale
within the flux concentration (`Wilson depression'). The relation
between the displacement, $z_{0.1}-\langle z_{0.1}\rangle$, and
the field strength is shown in the second row of JPDFs in
Fig.~\ref{fig:u5_histograms}: the stronger the field, the larger
is the downward (negative) displacement of the $\tau_{5000}=0.1$
level, which can reach values up to 120~km. This has the
consequence that $|B_z|$ at the local level $\tau_{\rm 5000}=0.1$
has values of up to 1.5~kG, since the field strength increases
with geometrical depth (see the top right panel in
Fig.~\ref{fig:u5_histograms}).

The partial evacuation of vertical flux concentrations leads to
another observational signature. From the Eddington-Barbier
relation, we know that the emergent intensity is approximately
given by the source function at $\tau=1$. The panels in the third
row of Fig.~\ref{fig:u5_histograms} show JPDFs between $|B_z|$ at
$z=\langle z_{0.1}\rangle$ and the temperature at
$\tau_{5000}=1.0$. The temperature at optical depth unity is
higher in strong field regions, with the result that the
brightness of vertical flux tubes increases with $|B_z|$ (see
fourth row of Fig.~\ref{fig:u5_histograms}). The higher
temperature at $\tau_{5000}=1.0$ is due to the lateral radiative
heating of the interior of the flux concentration by the
surrounding `hot walls' \citep{Spruit:1976,
Deinzer:etal:1984,Grossmann-Doerth:etal:1994, Voegler:MURaM}.

\subsection{Emergence of strong magnetic field}
\label{subsec:emergence_of_strong_field} 
\begin{figure*}
\subfigure[Emergent intensity and horizontal flow
field]{\includegraphics[width=0.499\textwidth]{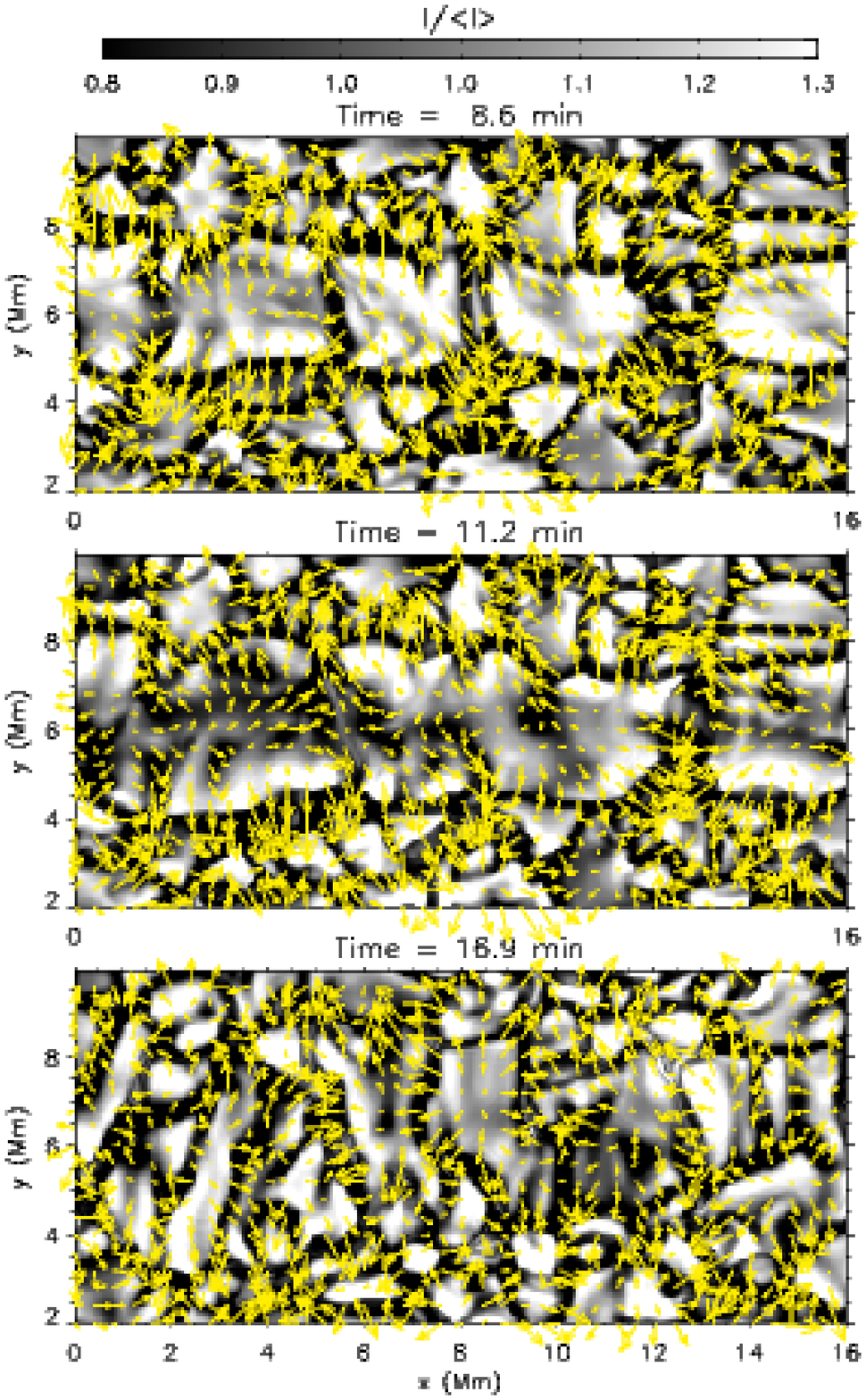}\label{fig:u1_intensity_sequence}}
\subfigure[Corresponding vector
magnetograms]{\includegraphics[width=0.499\textwidth]{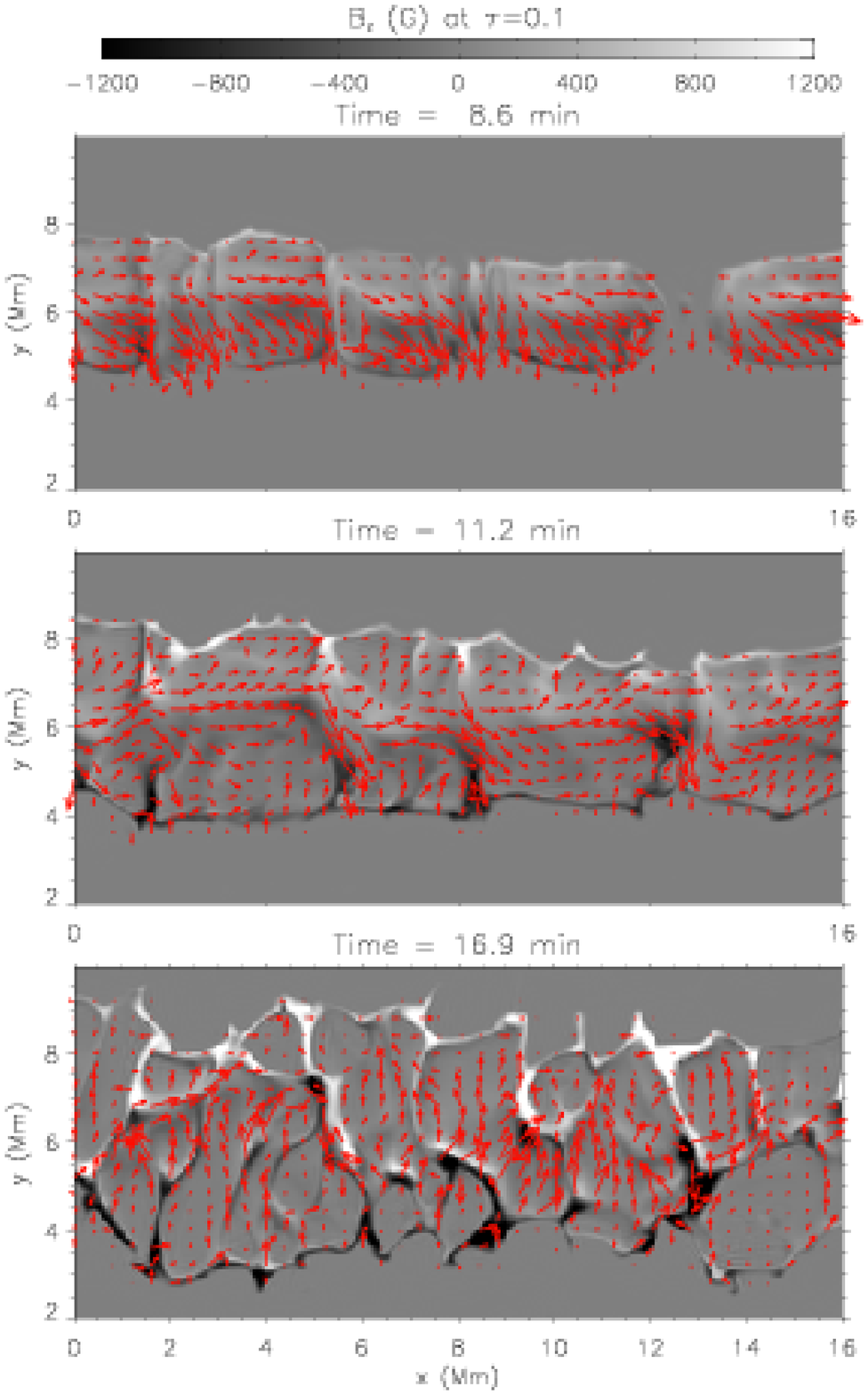}\label{fig:u1_magnetograms}}
\caption{Flux emergence event in run U1.~\emph{Left}: an arrow
with a length of one grid-spacing on the vector plot corresponds
to a horizontal flow speed of $v=3$~\kms.~\emph{Right}: an arrow
with a length of one grid-spacing on the vector plot corresponds
to a horizontal field strength of $|B|=400$ G. Both the velocity
and magnetic fields shown are evaluated at $\tau_{5000}=0.1$. The
granulation pattern is temporarily disturbed by the emerging
magnetic field and a transient darkening along the length of the
domain marks the site of emergence. An accompanying mpeg animation
is available at
http://www.mps.mpg.de/homes/cheung/U1emergence.mpg.}
\end{figure*}

In this section, we
discuss the observational signatures of the emerging flux tube
with $B_0=8500$ G (run U1), which is strong enough to resist being
passively deformed and advected by the convective flow.
Fig.~\ref{fig:u1_intensity_sequence} gives a time sequence of
continuum intensity maps throughout the emergence event (grey
scale). The vector overlays indicate the horizontal components of
the velocity field at $\tau_{5000}=0.1$.
Fig.~\ref{fig:u1_magnetograms} shows the corresponding maps of the
vector magnetic field.

We have seen that the emergence morphology of the field in run U6
is dictated by the convective motion of the near-surface layers.
In run U1 (where the magnetic field is several times stronger),
the convection still has an appreciable influence on the emergence
morphology. Comparison of the magnetic field and the intensity
maps at $t=8.6$ min clearly shows that magnetic flux emerges
preferentially within the interior of granules. On the other hand,
we find that the emerging magnetic structure also affects the
appearance of the local granulation pattern. At $t=8.6$~min (top
panel of Fig.~\ref{fig:u1_intensity_sequence}), there is an
alignment of granules along the direction of the tube axis. These
granules are relatively large, each covering an area of up to $9$
Mm$^2$. The disturbance of the granulation by the emerging
magnetic field is most prominent at $t=11.2$ min, showing a
darkening near the emergence site extending over the full
horizontal size of the simulation domain. This darkening is not
coherent along its entire length: there are bright patches on
either side of the darkening but their shapes are distinctly
different from those of normal granules. The expansion of the
emerging tube drives horizontal outflows away from the emergence
site, which are visible in the vector plots in the middle panel of
Fig.~\ref{fig:u1_intensity_sequence}. The magnetic field maps for
the same instant shows that much of the emerged flux already
resides in intergranular lanes.

At $t=16.9$ min, the darkening has largely vanished and a
granulation pattern more similar to the normal pattern has
re-established itself. However, these granules tend to be
elongated in the $y$-direction because of the horizontal outflows
driven by the expanding tube. In the corresponding magnetic field
map, we find that the emerged magnetic flux outlines the
boundaries of these new granules. The $y$-component of the
magnetic field is, on average, positive. This is just the opposite
to what we see in the magnetogram at $t=8.6$ min. This reversal of
the sign of $B_y$ within this time span is expected. In the
earlier snapshot, the $\tau_{\rm 5000}=0.1$ surface intersects
with the upper half of the flux tube. The twist of the field lines
within the tube is such that $B_y < 0$ in this part of the tube.
In the latter snapshot, the lower half of the tube has emerged at
the surface, and so the synthetic magnetogram shows field with
predominantly $B_y > 0$.

\subsubsection{The relation between field strength and zenith angle}
\citet*{Lites:EmergingFieldsVector} have presented observational
results for several emerging flux regions. They obtained vector
magnetograms by applying inversion methods to measurements of the
full Stokes vector. Based on their
observations,~\citet{Lites:EmergingFieldsVector} describe the
following scenario for flux emergence: magnetic flux emerges at
the surface in the form of horizontal field with field strengths
of $200-600$~G. After emergence, the field quickly migrates away
from the emergence site. In the process, the field becomes
vertical. Only then do the fields obtain field strengths exceeding
one kilogauss. The observational study
by~\citet{Kubo:EmergingFluxRegion} also supports these findings.

\begin{figure*}
\centering
\includegraphics[width=0.8\textwidth]{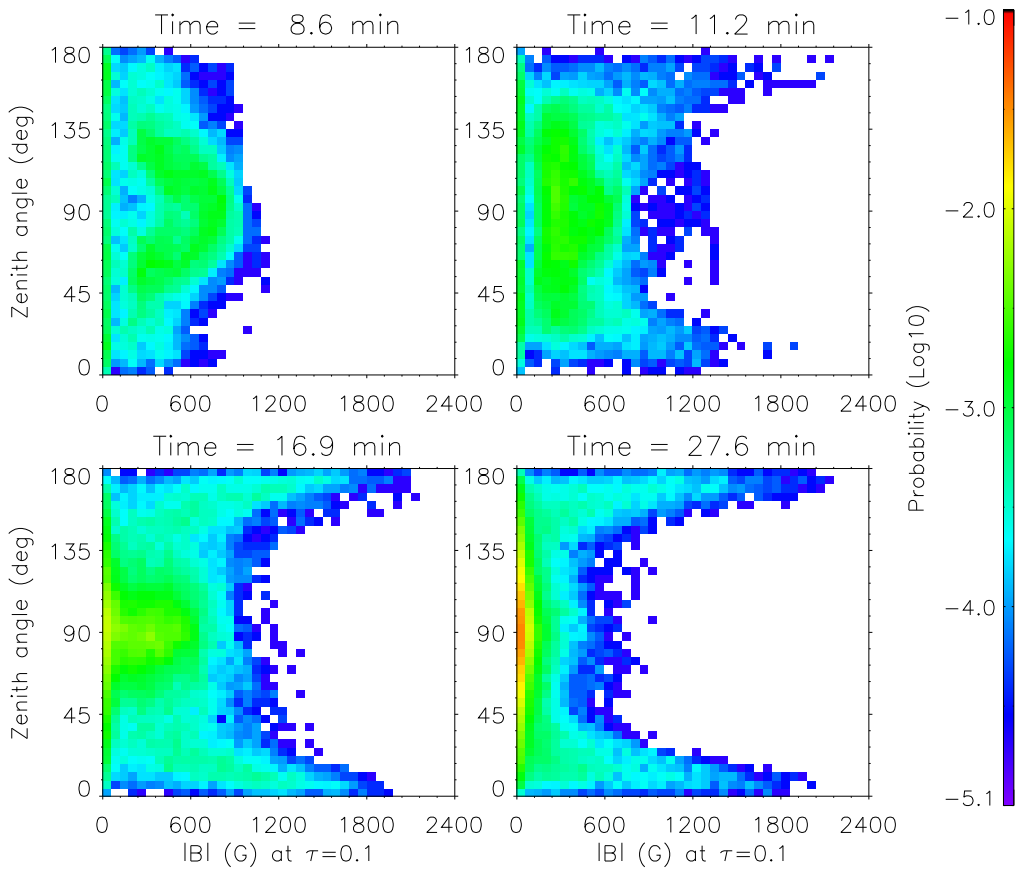}
\caption{Joint probability distribution functions (JPDFs) of the
zenith angle of the magnetic field vector versus the absolute
field strength $|B|$ in run U1. Both quantities have been
evaluated at the $\tau_{\rm 5000}=0.1$ level. A zenith angle of
$90^\circ$ corresponds to a purely horizontal field. The four
panels show the JPDFs at four different times. At $t=8.6$ min, the
flux tube is emerging at the surface and the strongest fields are
predominantly horizontal. Kilogauss fields develop only later and
are almost exclusively vertical at 27.6~min.}
\label{fig:u1_inclination}
\end{figure*}

Figure~\ref{fig:u1_inclination} shows JPDFs of the zenith angle of
the magnetic field vector versus the absolute field strength from
simulation run U1. At $t=8.6$~min, the flux tube is just emerging
at the surface. The JPDF at this time reflects the fact that the
flux tube is still coherent. The horizontal fields are stronger
than the vertical fields. Although a tiny fraction of the
horizontal fields have kilogauss field strengths, most are
confined within the range $400 \leq |B| \leq 1000$ G. Just five
minutes later, at $t=11.2$ min, the distribution in the JPDF looks
very different. The appearance of two `horns' in the JPDF near
$\gamma=0^\circ$ and $\gamma=180^\circ$ indicates the effect of
the granular flow on the emerged field. As the emerged field is
advected to the boundaries of granules, it also becomes
predominantly vertical. As the simulation progresses in time, the
two horns in the JPDF become more distinct. At $t=27.6$ min,
nearly all the field with $|B|\geq 600$ G is vertical. There is
actually an abundance of horizontal fields, but these are confined
to strengths of less than $600$ G. The shape of the JPDF at this
time is similar to the scatter plots of zenith angle vs. $|B|$
obtained by~\citet{Lites:EmergingFieldsVector} and
by~\citet{Kubo:EmergingFluxRegion}.

\subsubsection{The dark lane}
\label{subsubsec:dark_lane}

\begin{figure*}
\centering
\includegraphics[width=0.8\textwidth]{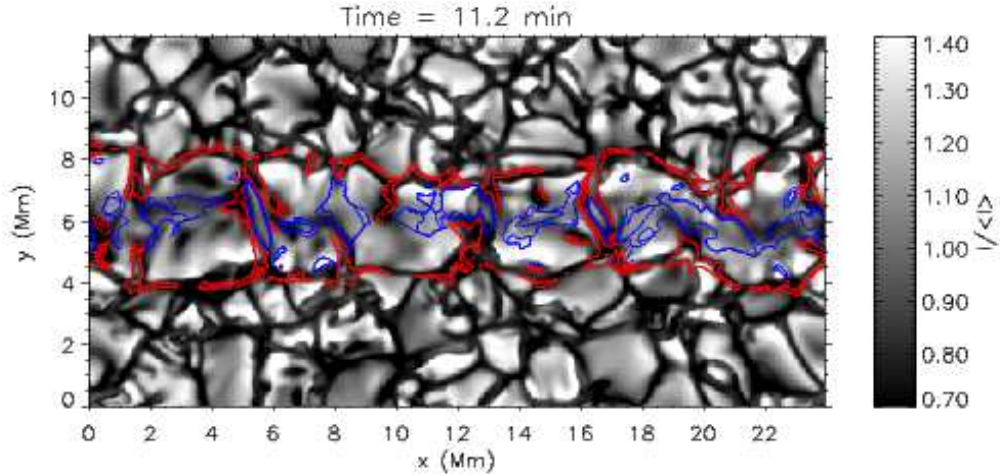}
\caption{Dark lane ($y\sim 6$ Mm) in run U1 ( $t=11.2$~min). The
grey shading represents the normalized emergent continuum
intensity. The blue contours indicate upflows with $v_z=0.5$~\kms
and 1~\kms, respectively, while the red contours correspond to
downflows of the same magnitude ($v_z<0$). The dark lane is
roughly coincident with upflowing material.}
\label{fig:u1_dark_lane}
\end{figure*}

\begin{figure}
\centering
\includegraphics[width=0.6\textwidth]{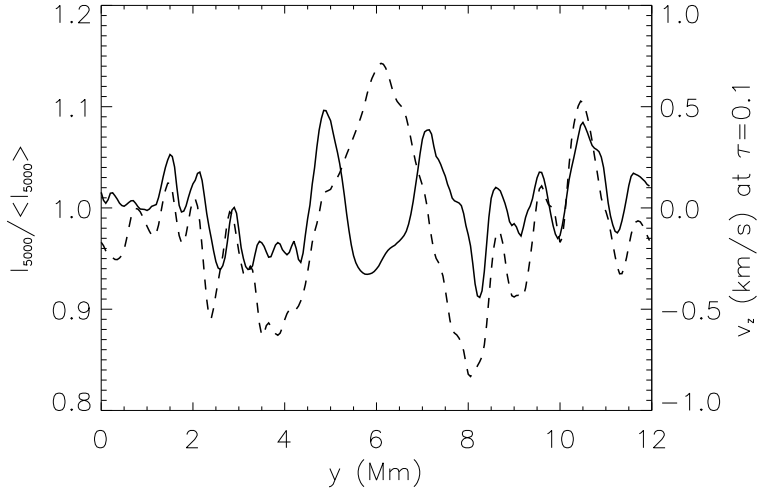}
\caption{Average profiles of the normalized continuum emergent
intensity (solid line, left axis) and vertical velocity at
$\tau_{5000} = 0.1$ (dashed line, right axis) across the dark lane
at $t=11.2$ min. The curves represent averages over the
$x$-direction.} \label{fig:u1_dark_lane_ic_vz}
\end{figure}

The physical nature of the extended dark lane (see central panel
of Fig.~\ref{fig:u1_intensity_sequence}) is distinctly different
from that of normal intergranular lanes, which are associated with
cold downflowing material. Instead, the dark lane that results
from the emergence of the flux tube in run U1 is associated
with~\emph{upflowing}~material for a substantial fraction of its
lifetime (several minutes). Fig.~\ref{fig:u1_dark_lane} shows the
normalized emergent continuum intensity at $t=11.2$ min (same as
the central panel in Fig.~\ref{fig:u5_coalescence_event}).
Overplotted on the grey scale image are contours of the vertical
component of the velocity at $\tau_{\rm 5000}=0.1$ for magnetic
regions ($|B| \geq 400$ G). The blue contours indicate upflows
while the red contours correspond to downflows. There is a
substantial overlap of the blue contours with regions on the
surface comprising the dark lane. The correlation between upflows
and low emergent intensity in the dark lane is also illustrated in
Fig.~\ref{fig:u1_dark_lane_ic_vz}, which shows profiles of
emergent continuum intensity and vertical velocity at $\tau_{\rm
5000}=0.1$ across the dark lane (in the $y$-direction). The
plotted values are averages taken along the $x$-direction at time
$t=11.2$ min. Away from the emergence site, we have the usual
intensity-velocity correlation of granules and intergranular
lanes. However, within the emergence site (around $y\simeq 6$ Mm),
the darkening ($I_{5000}/\langle I_{5000}\rangle\approx 0.93$) is
co-spatial with upflowing material ($v_z=0.5-1$~\kms).

The lifetime of the extended dark lane is about $10$ minutes,
which is comparable to the timescale of granulation. During the
first few minutes, the regions of lower brightness are associated
with upflowing material with rise speeds of up to $1$~\kms. This
is substantially lower than the typical rise speed ($v\approx
2-4$~\kms) of magnetic material when it is just emerging at the
visible surface. In the last few minutes of the dark lane's
lifetime, the dark lane splits up into spatially separated, dark
elongations with length of about $2$ Mm (see intensity image at
$t=16.9$ min in Fig.~\ref{fig:u1_intensity_sequence}). The
vertical velocities of these dark elongations are negative, i.e.,
they have developed into downflows.
The presence of the dark lane depends on the initial field
strength as well as the initial twist carried by the tube. When
the initial flux tube is sufficiently weak that its evolution is
dominated by the drag force (as is the case with run U6), a dark
lane does not appear: the magnetic flux is passively carried
upward by the convective flow, which shows the normal
intensity-velocity correlation. In order for the dark lane to
appear, the emerging field must be strong enough that its
evolution is not completely controlled by the convective flow.
However, this is only a necessary criterion. The amount of initial
twist is also important. We have carried out simulation runs with
the same initial conditions for the flux tube as in run U1,
varying only the amount of the initial twist. We found that for a
flux tube with twist parameter $\lambda=0$, there is no transient
dark lane marking the emergence site of the tube. For a flux tube
with a moderate amount of twist, $\lambda=0.25$, we find the
transient appearance of some dark patches within the granules
where the emergence occurs. These patches, however, do not exhibit
a clear alignment or coherence as is the case in run U6 (where
$\lambda=0.5$).
The following explanation accounts for the properties we have
presented above: When a flux tube (with a flux of, say,
$\Phi_0=10^{19}$ Mx) reaches the surface, it has a rise speed of
$2-4$~\kms. As demonstrated by previous numerical MHD simulations
of the rise of twisted flux tubes through an initially static
atmosphere~\citep*[see, for
example,][]{Moreno-InsertisEmonet:RiseOfTwistedTubes,
EmonetMoreno-Insertis:PhysicsOfTwistedFluxTubes,
Cheung:MovingMagneticFluxTubes}, the amount of twist in the tube
determines whether it can remain a coherent structure (large
$\lambda$) or break up into vortex fragments (small $\lambda$). As
the flux tube emerges and cools by radiation, it loses buoyancy.
Nevertheless, the emerged magnetic material may still overshoot
provided that the flux tube is sufficiently twisted, so that its
core remains largely intact and maintains its magnetic buoyancy.
As the emerged magnetic material of the tube overshoots, its
upwards motion is braked by the stable stratification. This is why
the upward velocities in Fig.~\ref{fig:u1_dark_lane} are limited
to $0.5-1$~\kms~within the dark lane. The strong cooling of the
magnetic material within the core of the tube at optical depth
unity accounts for the lane's darkness. The arrangement of the
cool and dense material in an extended lane configuration is not
stable to perturbations introduced by granular dynamics. Within a
few minutes, the extended dark lane breaks up into shorter
elongations with lengths comparable to those of intergranular
lanes. The buoyancy loss of the material by radiation and
stratification either leads to the birth of new downflows, or
feeds pre-existing downflows in the intergranular network. The
influence of the granular dynamics on its evolution explains why
the dark lane has a lifetime comparable to the granulation
timescale.

Transient appearance of dark alignments in emerging flux regions
have been reported by a number of observational
 studies~\citep*{BrayLoughhead:SolarGranulation,
 BrantsSteenbeek:EFR,Zwaan:EmergenceOfMagneticFlux,
 StrousZwaan:SmallScaleStructure}. For example,
 \citet{StrousZwaan:SmallScaleStructure} have carried out a statistical
 analysis of an emerging flux region and found that transient
 darkenings in the continuum as well as in the cores of photospheric
 spectral lines, lasting about $10$ minutes, are robust signatures of
 emergence events. They note that the typical extension of the
 darkenings is about $2$ Mm and that they are aligned roughly parallel
 to the direction connecting the two developing polarities of the
 corresponding active region. \citet{StrousZwaan:SmallScaleStructure}
 also report that the darkenings are associated with upflows of about
 $0.5$~\kms. These properties are reproduced by our simulation results
 from run U6. In addition,~\citet{StrousZwaan:SmallScaleStructure}~also
 report that the darkenings, on average, rotate counterclockwise at
 about 0.5$^\circ$ min$^{-1}$. In run U6, we do find that some of the
 dark elongations rotate during the breakup of the dark lane. However,
 we do not find any systematic direction in which the rotation
 occurs.

\section{Discussion}
\label{sec:summary} The magnetic flux emergence simulations
presented in this paper are `realistic' in two senses. Firstly,
the simulations take into account the effects of non-local
radiative energy exchange, partial ionization and
magneto-convection, all of which are important for a proper
treatment of the problem. Secondly, the simulations yield
observational signatures of magnetic flux emergence that are in
qualitative and quantitative agreement with observations of
emerging flux.

Our simulation results are generally consistent with the findings
of~\citet*{Fan:3DTubeConvection}, extending their validity to the
near-surface layers, where the anelastic approximation used by
these authors is no longer tenable. However, Mach numbers reaching
$M\sim O(0.1) - O(1)$ and the steep decrease of pressure and
density with height limit the range of possible ratios of initial
and equipartition field strengths, $B_0/B_{\rm eq}$. As a
consequence, a considerable influence of the convective flows on
the dynamics of the rising flux tube and on the flux emergence
process is unavoidable.

Convection influences the evolution of emerging magnetic fields
both before and after they appear at the photosphere. In the
subphotospheric layers, the convective flow distorts a rising flux
tube and imposes a systematic undulation along the tube (with a
typical wavelength $\sim 1-2$ Mm). Whereas upflows aid the rise of
certain segments of the tube, downflows suppress the rise of other
segments. Our work shows that the undulatory nature of an emerging
magnetic tube~\citep{Pariat:ResistiveEmergence} is a consequence
of its interaction with granular convection.

The amount of distortion and deformation suffered by a magnetic
flux tube interacting with convective flows depends on its initial
properties, most notably its buoyancy and twist. Magnetic flux
tubes with less longitudinal flux (say, $10^{18}$ Mx) are not
sufficiently buoyant to rise coherently against the convective
flows. The emergence events associated with these smaller and
weaker flux tubes are inconspicuous, in the sense that the
granulation pattern in the quiet Sun is not disturbed (see
Section~\ref{subsec:quiescent_flux_emergence}). Their small
spatial scales ($\sim 1$ Mm) and short temporal scales ($5$ min)
make their observations difficult. Although detections of such
events have been
reported~\citep{DePontieu:Small-scaleEmergingFlux}, it is unknown
how much flux emerges at the surface in this form. High-cadence,
high-resolution observations from the Solar Optical Telescope
onboard
Hinode~\citep[Solar-B,][]{Ichimoto:SOT,Tarbell:FocalPlanePackage}
will allow us to study such events in more detail then ever
before.

In Section~\ref{subsubsec:surface_evolution}, we provided an
example of the secondary emergence of a bipole. This case (with a
flux of $10^{18}$ Mx in each polarity) is peculiar for two
reasons. Firstly, the bipole emerged rather distant from the main
emergence site. Secondly, it emerged several granulation time
scales after the initial appearance of flux at the surface. This
secondary emergence event is the result of the recirculation and
overturning of material in the near-surface layers of the
convection zone. It suggests to us that part of the small-scale
flux emergence events on the solar surface could result from
recirculation of material in the convection zone.

The expulsion of emerged magnetic flux from the granule interiors
to the intergranular network on the order of the granulation time
scale ($5-10$ min) is a common feature of all our flux emergence
simulations. Having said this, not all emergence events appear the
same. For example, the observational signatures of the emergence
of tubes with $10^{19}$ Mx flux are clearly different from those
of tubes with $10^{18}$ Mx. Although they still suffer distortion
and deformation by the convective flow, such bigger flux tubes are
able to rise against downflows and eventually emerge at the
surface. During emergence, the strong horizontal expansion of
rising fluid can disturbs the granulation pattern. If the flux
tube is sufficiently twisted (say, $\lambda=0.5$), the emergence
event is accompanied by the transient appearance of an extended
darkening in the direction of the tube axis. Such features have
been reported in observations of emerging flux regions. Analysis
of our simulations results reveals that they are associated with
emerged, cooled magnetic material overshooting into the
photosphere.
\\ \\
\noindent \textbf{Acknowledgements} \\The authors thank the
anonymous referee for helping to substantially improve the
presentation of this paper. MCMC wishes to thank all the friends
and colleagues at the IAC for their hospitality during his
extended stays in Tenerife as a visiting graduate student. In
particular, MCMC is deeply grateful to have known the late Juan
Luis Medina Trujillo, who will always be remembered for his
sincerity, his love of nature and his zest for life.

\end{document}